\documentclass[reprint, amsfonts, amssymb, amsmath, aps, prapplied, floatfix, twoside, superscriptaddress, longbibliography]{revtex4-1}

\usepackage{graphicx}
\usepackage{dcolumn}
\usepackage{bm}
\usepackage{caption}
\usepackage{array}
\usepackage[dvipsnames]{xcolor}
\usepackage{hyperref, cleveref}
\renewcommand{\in}{\mathrm{in}}
\newcommand{\out}{\mathrm{out}}

\newcommand{\pump}{\mathrm{pump}}
\newcommand{\leak}{\mathrm{leak}}
\newcommand{\sys}{\mathrm{sys}}
\renewcommand{\Im}{\mathrm{Im}}
\renewcommand{\Re}{\mathrm{Re}}

\renewcommand{\exp}[1]{e^{#1}}
\newcommand{\J}{\mathrm{J}}
\newcommand{\dB}{\mathrm{dB}}
\renewcommand{\(}{\left(}
\renewcommand{\)}{\right)}
\newcommand{\abs}[1]{\left| #1 \right|}

\newcommand{\Dif}[2]{\frac{\mathrm{d} {#1}}{\mathrm{d} {#2}}}

\newcommand{\eq}[1]{equation (\ref{#1})}
\newcommand{\fig}[1]{Figure.\ref{#1}}

\begin{document}

\title{Optimizing the pump coupling for a three-wave mixing Josephson parametric amplifier}

\author{Wei Dai}
    \email{wei.dai.wd279@yale.edu}
    \affiliation{Department of Applied Physics, Yale University, New Haven, CT 06520, USA}
\author{Gangqiang Liu}
    \affiliation{Department of Applied Physics, Yale University, New Haven, CT 06520, USA}
\author{Vidul Joshi}
    \affiliation{Department of Applied Physics, Yale University, New Haven, CT 06520, USA}
\author{Alessandro Miano}
    \affiliation{Department of Applied Physics, Yale University, New Haven, CT 06520, USA}
\author{Volodymyr Sivak}
    \affiliation{Department of Applied Physics, Yale University, New Haven, CT 06520, USA}
\author{Shyam Shankar}
    \affiliation{Department of Applied Physics, Yale University, New Haven, CT 06520, USA}
    \affiliation{Chandra Department of Electrical and Computer Engineering, University of Texas at Austin, Austin, TX 78712, USA}
\author{Michel H. Devoret}
    \email{michel.devoret@yale.edu}
    \affiliation{Department of Applied Physics, Yale University, New Haven, CT 06520, USA}

\date{\today}

\begin{abstract}
Josephson parametric amplifiers (JPAs) typically require rf pump power that is several orders of magnitude stronger than the maximum signal power they can handle. The low power efficiency and strong pump leakage towards signal circuitry could be critical concerns in application. In this work, we discuss how to optimize the pump coupling scheme for a three-wave mixing JPA by employing microwave filtering techniques, with the goal of maximizing the pump power efficiency and minimize pump leakage without sacrificing other properties of interest. We implement the corresponding filter design in a SNAIL-based JPA and demonstrate more than three orders of magnitude improvement in both power efficiency and pump leakage suppression compared to a similar device with conventional capacitive coupling, while maintaining state-of-the-art dynamic range and near-quantum-limited noise performance. Furthermore, we show experimentally that the filter-coupled JPA is more robust against noise input from the pump port, exhibiting no significant change in added noise performance with up to 4 K of effective noise temperature at the pump port. 
\end{abstract}

\maketitle

\begin{section}{Introduction}

Josephson parametric amplifiers (JPAs)~\cite{JPA_review} have become critical tools in quantum measurements in which the information is carried by a few microwave photons. These amplifiers provide large gain with near-quantum-limited added noise, enabling highly efficient microwave measurements in single-shot qubit readout~\cite{Jeffrey_2014_FastReadout}, electron spin resonance detection~\cite{Bienfait_2016_ESR}, and axion search~\cite{Brubaker_2017_Axion}. Recent research has focused on improving JPA bandwidth and compression power~\cite{Broadband_JPA,Eichler2014,Saturation_Gang,SPA,Naaman2019,Saturation_Planat,kaufman_Chebyshev2023,kaufman_SimpleHighSaturation2024}, which is necessary for multiplexed qubit readout~\cite{Kundu2019,whiteReadoutQuantumProcessor2023} in large-scale quantum processors;  
however, one operational challenge in scaling up superconducting quantum processors is the heat load added to the cryostat by the microwave drives applied to the JPAs, commonly referred to as the `pump'.

To achieve quantum-limited noise performance with a JPA, the pump is typically delivered to the JPA through a heavily attenuated transmission line to reduce thermal noise. 
In addition, the pump line is weakly coupled to the JPA, usually with a small capacitance or mutual inductance, to minimize the signal photon loss. 
As a result of these choices, the pump power required by the JPA is usually several orders of magnitude higher than the signal power. 
For instance, a state-of-the-art three-wave-mixing JPA dissipates around 0.01 to 1~\textmu W of power on the base plate of a dilution refrigerator, whose cooling power is typically 10 to 100~\textmu W at 20~mK~\cite{Kerr_free}.
This is a practical limiting factor in operating a large array of JPAs at millikelvin temperatures.

In addition, the strong pump power incident on the JPA also poses challenges for protecting qubits against unwanted back-action~\cite{kaufman_SimpleHighSaturation2024}. 
Since the coupling strength on the signal port is much higher than that on the pump port, the intra-amplifier pump photons predominantly exit the amplifier through the signal port and propagate toward the signal source. 
Multiple nonreciprocal components (other than the circulator shown in Fig.~\ref{fig:schematic}) or filters are usually required to achieve sufficient isolation of the pump leakage from the signal source. 
These elements reduce readout efficiency due to their insertion loss, and they can emit thermal photons that dephase the qubit due to heating by the pump leakage. 

\begin{figure}[htb]
\includegraphics{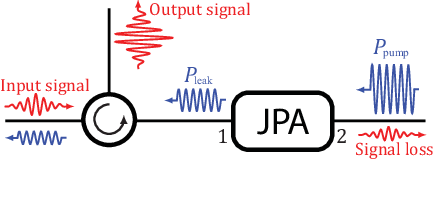}
\caption
{\label{fig:schematic}
Minimal schematic describing a reflection JPA as the first stage of the amplifier chain, illustrating the problems addressed in this article. Red arrows indicate information-carrying signals and blue arrows indicate pump waveforms. There is generally a trade-off between signal loss and the required pump power at port 2. Furthermore, one circulator at port 1 (for signal directionality) is not sufficient to isolate pump leakage from the input circuitry. 
}
\end{figure}

Reducing the pump power needed to operate JPAs is therefore crucial for two reasons: reducing the heat load and improving both the readout performance and qubit coherence. 
These issues become increasingly pressing because broadband, higher-saturation-power JPAs often require elevated intra-amplifier pump power ~\cite{Saturation_Planat,JPC_Hamiltonian_control,JAMPA,kaufman_SimpleHighSaturation2024} to operate. 
This therefore points to a clear design objective to increase pump-power coupling at the pump port while further suppressing pump leakage at the signal port.

JPAs with pump tones far detuned from the signal frequency offer the opportunity to address these issues by applying microwave filtering techniques. 
In this work, we demonstrate a near-30-dB reduction in both pump power and pump leakage in a three-wave-mixing SNAIL parametric amplifier (SPA) by upgrading its capacitively coupled ports with on-chip filters, while retaining its near-quantum-limited noise performance. 
These improvements would enable deployment of a large bank of parallel SPAs with minimal heat load added to the cryostat, facilitating the scaling up of superconducting quantum processors. 
The technique introduced in this work is also applicable to four-wave-mixing parametric amplifiers with sideband pumping~\cite{Double-pump}, and more generally to other parametric processes activated by off-resonant pumps. 

The remainder of this article is organized as follows: In Section II, we introduce \emph{power efficiency} as the metric for characterizing the efficiency of converting pump power to the output signal power of a JPA, and we formulate its optimization as a specific circuit synthesis problem. 
In Section III, we demonstrate a realization of a filter-coupled SPA that fulfills the optimization goals. 
In Section IV, we compare the filter-coupled SPA with a control device with conventional capacitive coupling, demonstrating improvements in both power efficiency and pump-leakage suppression. 
In Section V, we show the improved noise rejection at the signal frequency on the pump port of the filter-coupled SPA, which would allow the amplifier to achieve near-quantum-limited noise performance with reduced attenuation on the pump line and further decrease its heat load to the cryostat. 

\end{section}
\begin{section}{Optimization of JPA power efficiency}

A parametric amplifier converts power from its pump into an output signal. 
The figure of merit that characterizes the efficiency of this process is the \emph{power efficiency} $\eta_p$, which is defined as the ratio between the output signal power at 1-dB gain compression point ($P^\out_{1\dB}$) and the corresponding pump power ($P_\pump$): 
\begin{equation}\label{eq:efficiency}
	\eta_p = \frac{P^\out_{1\dB}}{P_\pump}.
\end{equation}

JPAs typically have power efficiencies ranging from $10^{-2}$ to $10^{-6}$. 
As reviewed in Ref.~\cite{Kerr_free}, the power efficiency of three-wave-mixing JPAs is generally orders of magnitude lower than that of their four-wave-mixing counterparts. 
In the current-pumping scheme~\cite{Saturation_Gang,SPA,Saturation_Planat,JAMPA}, the pump of a three-wave-mixing JPA is handicapped in coupling strength due to a large frequency detuning from resonance. 
In contrast, in the flux-pumping scheme, the pump coupling is insensitive to detuning since a differential mode is driven, which is also advantageous in terms of suppressing pump leakage~\cite{Yamamoto2008};
However, it is challenging to engineer a strong mutual-inductive coupling. 
As a result, the reported power efficiencies for flux-pumped JPAs are typically at the $10^{-5}$ level~\cite{Simoen2015,Zhou2014,Elo2019,Mutus2013}, although there is ongoing research on improving these values ~\cite{Urade_MM2021}. In this section, we formulate the goals and constraints for optimizing the power efficiency of current-pumped JPAs within a network synthesis framework.

We consider a three-wave-mixing JPA operating in the phase-preserving mode, in which the wave components at pump frequency $\omega_p$, signal frequency $\omega_s$, and idler frequency $\omega_i = \omega_p - \omega_s$ are coupled via the third-order nonlinearity. 
We focus on a current-pumped JPA whose nonlinearity originates from a \emph{lumped} Josephson dipole, namely a two-terminal Josephson element. 
As illustrated in Fig.~\ref{fig:circuit}(a), we view the linear embedding circuit of the Josephson dipole in the JPA as a microwave network with two external ports connected to the signal and pump sources respectively, and one internal port loaded by the Josephson dipole. 
Separate signal-flow analyses are performed at the pump and signal frequencies to formulate the power efficiency and amplification performance in terms of the circuit parameters. 

\begin{figure}[htb]
\includegraphics[width = 8.6cm]{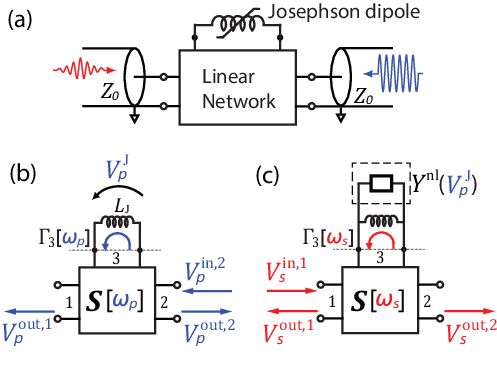}
\caption
{\label{fig:circuit} 
Modeling the JPA as a scattering system. (a) The JPA consists of a lumped Josephson dipole embedded in a linear network, which has two external ports and one internal port. The external ports — port 1 and port 2 — connect to the signal and pump source, respectively. The internal port — port 3 — is loaded by the Josephson dipole. (b) The pump at frequency $\omega_p$ is applied from port 2, creating a voltage phasor $V_p^{\J}$ across the Josephson dipole. (c) At signal frequency $\omega_s$, the Josephson dipole produces an additive admittance arising from pumping. The reflection gain for the signal is measured from port 1. }
\end{figure}

We decompose the power efficiency into two meaningful factors by rewriting the definition in Eq.~\eqref{eq:efficiency} as
\begin{equation}\label{eq:eta}
\eta_p = \frac{P_{1\mathrm{dB}}^\out}{P^{\J}_{p}} \cdot \frac{P^{\J}_{p}}{P_\pump}
\end{equation}
where $P^{\J}_{p} = \abs{V^{\J}_{p}}^2/\omega_p L_{\J}$ is the steady-state pump power resulting from the rms voltage $\abs{V_P^{\J}}$ across the Josephson dipole. 
The first factor in Eq.~(\ref{eq:eta}) is the ratio between the output signal power at 1-dB gain compression point and the pump power incident on the Josephson dipole. We will refer to this factor as the \emph{nonlinear mixing efficiency} $\eta^{\mathrm{nl}}_p$. 
The second factor in Eq.~(\ref{eq:eta}) is the ratio between the pump power incident on the Josephson dipole and the power $P_\pump = |V^{\in,2}_{p}|^2/Z_0$ incident on the device pump port (port 2), which we will refer to as the \emph{pump coupling efficiency} $\eta^{\mathrm{c}}_p$. Hence $\eta_p = \eta^{\mathrm{nl}}_p \eta^{\mathrm{c}}_p$. 
Under the stiff-pump approximation~\cite{JPA_review}, these two factors are independent of each other. 

The power efficiency of a JPA can be improved by increasing the nonlinear mixing efficiency $\eta^{\mathrm{nl}}_p$ and/or the pump coupling efficiency $\eta^{\mathrm{c}}_p$. 
Improving $\eta_p$ by increasing $\eta^{\mathrm{nl}}_p$ has been demonstrated in experiments. 
For example, by operating a three-wave-mixing JPA at its `Kerr-free' point, where the fourth-order nonlinearity is minimized, the $P^{\out}_{1\dB}$ value can be improved by more than 10~dB under similar pump power ~\cite{Kerr_free}. 
Recently, a general method for engineering higher-order nonlinearities of Josephson elements for better JPA power efficiency has been proposed ~\cite{hougland2024pumpefficient}. 
However, it is important to note that obtaining a $\eta^{\mathrm{nl}}_p$ value close to unity leads to the effect of pump depletion~\cite{JPA_review, Schackert2013}, causing the device to deviate from the linear amplifier regime. 
More investigation is required for a JPA operated in the pump-depletion regime. 

In this work, we focus on improving $\eta_p$ by increasing the pump coupling efficiency. 
The goal is to establish the desired $P^{\J}_{p}$ across the Josephson dipole with minimal pump power $P_\pump$. 
This goal amounts to maximizing:
\begin{equation}\label{eq:xipump}
\eta_p^{\mathrm{c}} = \abs{\frac{V_p^{\J}}{V^{\rm{in},2}_p}}^2 \frac{Z_0}{\omega_p L_{\J}}
\end{equation}
for the circuit shown in Fig.~\ref{fig:circuit}(b), where the nonlinear Josephson dipole is approximated as a linear inductance $L_{\J}$ at the pump frequency.
Optimizing the pump coupling is thus formulated as a linear network synthesis problem. 
Notably, in contrast to the nonlinear mixing efficiency $\eta^{\mathrm{nl}}_p$, engineering $\eta^{\mathrm{c}}_p$ to its theoretical upper limit of unity does not lead to pump depletion. 



Furthermore, increasing $\eta^{\mathrm{c}}_p$ also helps suppress the pump leakage from the signal port (port 1). 
Assuming no loss in the circuit, conservation of the pump power requires: 
\begin{equation}\label{eq:leakage}
\frac{\abs{V^{\out,1}_{p}}^2}{Z_0} + \frac{\abs{V^{\out,2}_{p}}^2}{Z_0} = \frac{\abs{V^{\in,2}_{p}}^2}{Z_0} = \frac{P^{\J}_{p}}{\eta_p^{\mathrm{c}}}
\end{equation}
in which $P_\leak = |V^{\out,1}_{p}|^2/Z_0$ is the pump leakage power. 
Therefore, the pump leakage power scales as and is upper bounded by $P^{\J}_{p}/\eta_p^{\mathrm{c}}$, and improving $\eta_p^{\mathrm{c}}$ will generally reduce the pump leakage. 
We will discuss further suppression of pump leakage in Secs. III and IV by filtering at the signal port. 

The maximization of Eq.~\eqref{eq:xipump} is subject to constraints required for the circuit to function as a resonant amplifier. 
First, a standing wave eigenmode must exist in the circuit. This requirement is captured by the \emph{Laplace-domain} equation 
\begin{equation}\label{eq:resonance}
Z^\mathrm{Th}(s) + s L_{\J} = 0
\end{equation}
where $Z^\mathrm{Th} = Z_0 (1 + S_{33} )/(1 - S_{33})$ is the Thevenin impedance viewed by the Josephson dipole. 
The solution to Eq.~(\ref{eq:resonance}), $s_0= j \omega_0 - \frac{\kappa_0}{2}$, provides the resonance frequency $\omega_0$ and external coupling rate $\kappa_0$ for the electromagnetic eigenmode of the system. 
Eq.~(\ref{eq:resonance}) and its reformulation in the Fourier domain are discussed in Appendix C. 

Second, signal amplification requires the reflection coefficient from the Josephson dipole to have a magnitude greater than unity. 
Under the stiff-pump approximation, the response of the Josephson dipole at the signal frequency can be modeled as the inductance $L_{\J}$ in parallel with an effective admittance $Y^{\mathrm{nl}}(V^{\J}{p})$ arising from pumping, as illustrated in Fig.~\ref{fig:circuit}(c). 
This effective admittance representation resembles the `pumpistor' model developed for flux-pumped Josephson parametric devices~\cite{Pumpistor, Mutus2013,Sundqvist2014, naaman2021synthesis}, with details provided in Appendix D.
The reflection coefficient at the signal frequency is therefore given by:
\begin{equation}\label{eq}
\Gamma_3[\omega_s] = \frac{1 - Z_0 ( \frac{1}{j \omega_s L_{\J}} + Y^{\mathrm{nl}}(V^{\J}{p}) )}{1 + Z_0 (\frac{1}{j \omega_s L_{\J}} + Y^{\mathrm{nl}}(V^{\J}{p}) )}
\end{equation}
The requirement for amplification $\abs{\Gamma_3[\omega_s]} > 1$ is satisfied when $Y^{\mathrm{nl}}(V^{\J}{p})$ exhibits a negative real part. 

Lastly, for near-quantum-limited noise performance with high reflection gain, we require minimal signal loss via port 2, which demands $\abs{S_{23}[\omega_s]} \ll \abs{S_{13}[\omega_s]}$. 
In the conventional design with capacitively coupled ports, this condition is satisfied by having a large coupling capacitance at port 1 and a small coupling capacitance at port 2. However, this choice also leads to poor pump transmission from port 2, in conflict with our goal of maximizing pump coupling efficiency. 

The frequency separation between the pump and signal for the three-wave-mixing process provides an opportunity to satisfy these seemingly conflicting requirements. In the following sections, we demonstrate that the optimization goals at pump frequency $\omega_p$ and signal frequency $\omega_s$ can be achieved simultaneously by engineering the coupling network to be strongly frequency selective on a three-wave-mixing JPA. 

\end{section}
\begin{section}{Realization of a filter-coupled SNAIL parametric amplifier}

\begin{figure}[!ht]
\includegraphics[width = 8.6cm]{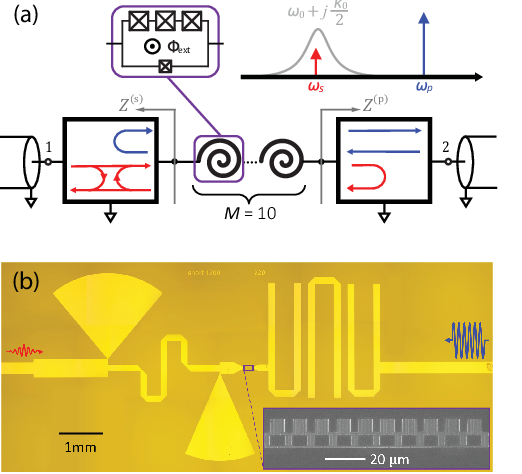}
\caption
{\label{fig:device}
Filter-coupled JPA in which the Josephson dipole is a SNAIL array.
(a) Schematic of a SNAIL parametric amplifier with two filters as coupling networks, with the signal-flow graphs representing the desired scattering properties at the signal (red) and pump (blue) frequencies respectively. The frequency landscape is sketched in the top right. (b) Optical image of a microstrip device designed to achieve the indicated scattering properties, with a scanning electron microscope image of the array of 10 SNAILs.
}
\end{figure}

In the last section, the goals and requirements for the circuit synthesis are formulated in terms of the scattering parameters of a 3-port linear network. 
However, it is technically challenging to directly synthesize this 3-port network in distributed-element circuit. Therefore, as illustrated in Fig.~\ref{fig:device}(a), we separate the design into a pair of 2-port networks — a pump-port network and a signal-port network — with an array of SNAILs~\cite{SNAIL} embedded in between, playing the role of the Josephson dipole. 
The pump-port network prevents transmission at the signal frequency $\omega_s$ while providing near-unity transmission at pump frequency $\omega_p$. 
Conversely, the signal-port network prevents transmission at $\omega_p$ while allowing partial transmission at $\omega_s$ with a rate $\kappa_0$, which is critical in forming the resonant mode of the amplifier. 
The signal-port network also suppresses pump leakage, in addition to the suppression originating from the improved pump coupling efficiency according to Eq.~(\ref{eq:leakage}).

Instead of the standard two-port microwave filter synthesis process based on scattering parameters~\cite{Pozar,matthaeiMicrowaveFiltersImpedanceMatching1980}, we formulated the design goals using the output impedances $Z^{(s)}$ and $Z^{(p)}$, viewed from each end of the Josephson dipole toward the signal and pump ports, respectively, as shown in Fig.~\ref{fig:device}(a). This allows the resonance condition for the concatenated circuit to be represented as an additional constraint on the sum of the two impedances $Z^{(s)} + Z^{(p)} = Z^{\mathrm{Th}}$ according to Eq.~(\ref{eq:resonance}). 
We discuss in Appendix B how to bridge between the output impedances and the scattering parameters, and how to formulate the resonance condition. 

For compatibility with our in-house fabrication and packaging process (Appendix A), we designed and implemented the amplifier in a microstrip architecture. 
We employed a compact three-stage hairpin bandpass filter (with the poles near $\omega_p$) for the pump-port network and a three-stage radial-stub lowpass filter for the signal-port network to achieve the desired frequency-selective properties. 
Specifically, the slow frequency roll-off property of the radial-stub lowpass filter allows us to position the amplifier mode within its transition band, facilitating control of $\kappa_0$. 

The synthesis process was conducted using AWR Microwave Office. 
We parametrized the hairpin and radial-stub structures based on our target signal (4.8-5.2~GHz) and pump (9.6-10.4~GHz) frequency bands. 
In addition, we aimed for a resonant mode at $\omega_0/2\pi \approx 5$~GHz with linewidth $\kappa_0/2\pi \approx 100$~MHz dominated by signal-port damping.
Formulating these requirements as constraints on the output impedances of the filter networks, we optimized the design parameters and verified the layout through electromagnetic simulations in AWR Axiem.

The output impedances $Z^{(s)}[\omega]$ and $Z^{(p)}[\omega]$ are evaluated assuming $50 \Omega$ matching at ports 1 and 2. 
Ensuring this matching condition experimentally across all relevant frequency bands is important for proper implementation, which requires attention, as pump frequencies often exceed the operational band of external microwave components. 

This filter-coupled SNAIL parametric amplifier, henceforth referred to as the F-SPA, was fabricated using a single-step electron beam lithography process on a silicon wafer as described in Appendix A. 
Fig.~\ref{fig:device}(b) presents an optical microscope image of the device, with an inset electron microscope image showing the array of 10 SNAILs, fabricated using a Dolan bridge process.

\end{section}
\begin{section}{F-SPA characterization and comparison with a conventional SPA}\label{Sec:pumping}

In this section, we report the characterization of the performance of the F-SPA described in the last section, and demonstrate its advantages over an SPA of conventional design~\cite{SPA,Kerr_free}. 

\begin{figure}[!hbt]
\includegraphics[width = 8.6cm]{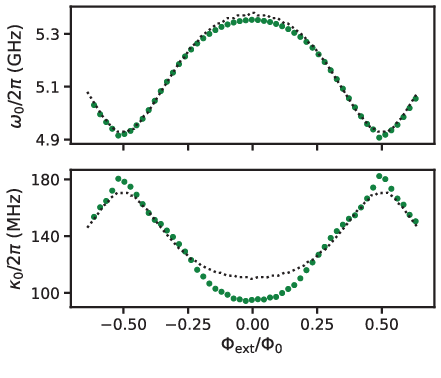}
\caption
{Resonance frequency $\omega_0$ and linewidth $\kappa_0$ of the device as functions of external flux, showing agreement between simulation (dashed lines) and measurement (points). 
}\label{fig:flux_sweep}
\end{figure}

We first measure the linear reflection from port 1 to extract the resonance frequency and linewidth of the F-SPA mode at different external flux bias values $\Phi_\mathrm{ext}$, as shown in Fig.~\ref{fig:flux_sweep}. 
The measured resonance frequency agrees well with that from the AWR Axiem simulation over the full flux tunable range of the device. 
The measured linewidth, however, is noticeably lower than that from the simulation near zero external flux. 
This deviation is likely caused by fabrication uncertainty, which shifts the pass band of the signal-port filter, and impedance mismatch on this port. 
Due to the lowpass nature of this radial-stub filter, the linewidth of the F-SPA mode is more sensitive to the frequency roll-off of the filter when in the high frequency part of its tunable range. 

At each flux bias point, from port 2, we then apply a pump tone at frequency $\omega_p = 2 \omega_0$ with the power that generates a small-signal 20-dB phase-preserving gain in reflection. 
Under this condition, we measure the pump leakage power $P_\leak$ (at port 1), output signal 1-dB compression power $P_{1\dB}^\out$ and the added noise at the frequency 100~kHz below $\omega_0$. 

\begin{figure*}[!htb]
\includegraphics[width=17.2cm]{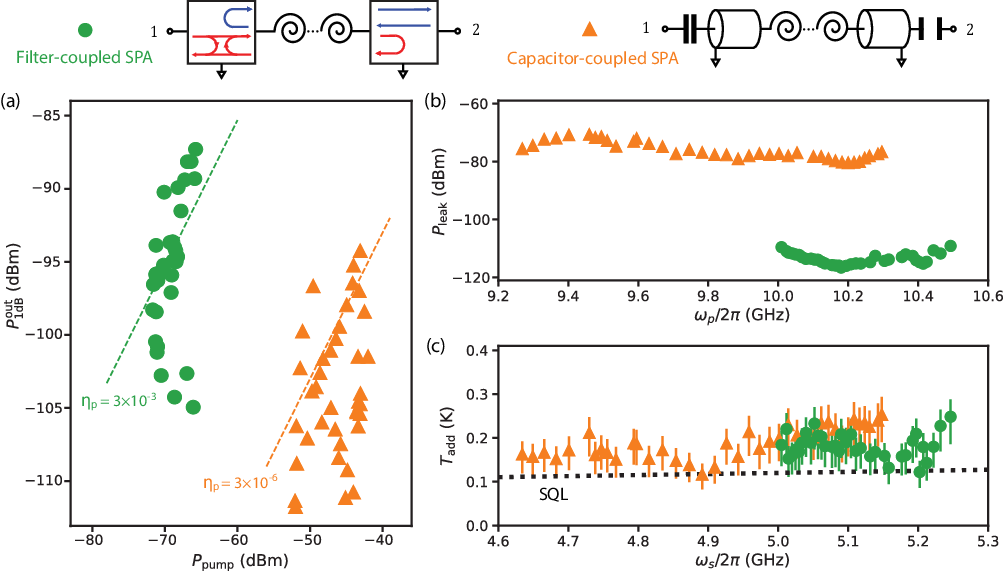}
\caption{
\label{fig:data} 
Characterization of the F-SPA (green dots) and C-SPA (orange triangles) with 20 dB reflection gain over their operating frequency ranges. (a) Output signal power at 1-dB compression points versus pump power at device pump port. Positions of the sloped lines indicates power efficiency $\eta_p$. (b) Pump leakage power measured at device signal port. (c) Added noise temperature $T_\mathrm{add}$, with a black dashed line indicating the standard quantum limit (SQL) $\hbar \omega / 2 k_B$. }
\end{figure*}

To demonstrate the advantage of the F-SPA, we compare its power efficiency, pump leakage, and noise performance with that of a control device using capacitive coupling at both ports~\cite{Kerr_free}. The latter consists of an array of SNAILs embedded at the center of a half-wave transmission line resonator, with a large coupling capacitance at the signal port and a small coupling capacitance at the pump port (see the schematics in the legend of Fig.~\ref{fig:data}). 
We referred to this device as the capacitor-coupled SPA or C-SPA.
 
For a fair comparison, the resonance frequency, linewidth and three-wave mixing strength of the C-SPA and the F-SPA are designed to be nominally identical. 
The resonance frequencies of both devices are designed to be between 4.5~GHz and 5.5~GHz such that their pump frequencies, between 9.0~GHz and 11.0~GHz, are within the pass band of the microwave components on the output line. 
Due to fabrication uncertainty, the linewidth of the C-SPA turned out in the range of 60 to 90~MHz, approximately a factor of two smaller than that of the F-SPA. 
This would result in about 6~dB lower pump power required for the same 20-dB gain.

We perform identical characterization measurements on the two amplifiers using the same experimental setup. 
As shown in Fig.~\ref{fig:data}(a), the F-SPA requires 20 to 25~dB less pump power than the C-SPA to generate 20~dB gain across the full operating range. 
The F-SPA shows higher $P_{1\mathrm{dB}}^\mathrm{out}$ due to its larger linewidth compared to the C-SPA. 
Overall, the F-SPA achieves an average power efficiency of $\eta_p = 3 \times 10^{-3}$, which is 30 dB better than the C-SPA and consistent with the 27-dB improvement predicted by our linear circuit simulation. 

The pump leakage power, as plotted in Fig.~\ref{fig:data}(b), is on average -113~dBm for the F-SPA and -77~dBm for the C-SPA. 
This dramatic 36-dB improvement in suppression of pump leakage on the F-SPA originates from the combination of the improved pump coupling efficiency and the reflectivity from the signal-port filter. 

It is worth pointing out that the pump for a typical three-wave-mixing JPA could fall outside the band of the circulators between the signal source and the JPA. 
The out-of-band reverse isolation of the circulators is generally not optimized. 
It is therefore of practical advantage to suppress pump leakage at the device level instead of relying on external components. 
 
The added noise temperatures (with respect to the input signal) of these two amplifiers are shown in Fig.~\ref{fig:data}(c). 
The added noise and output power of the amplifiers are calibrated using a shot-noise tunnel junction (SNTJ)~\cite{SNT_Malnou2024} as explained in Appendix A. 
Both devices have added noise that is at most twice the quantum limit for phase-preserving amplification~\cite{Cave}. 
The excess added noise is consistent for both devices, which is most likely due to residual thermal noise from the signal input line. 
Thermal noise from the pump line, however, does not contribute to added noise in these measurements as we will discuss in the next section. 

\end{section}
\begin{section}{Robustness against pump port thermal noise on the F-SPA}

The F-SPA requires significantly less pump power to operate compared to the C-SPA, leading to a reduced heat load to the dilution refrigerator from power dissipation on the pump line cryogenic attenuators. 
In this section, we investigate the effect of thermal noise from the pump line on these two devices. 
Remarkably, we demonstrate that the F-SPA maintains its near quantum-limited noise performance even when exposed to 4 K noise from the pump port. 
This characteristic affords an opportunity for further heat load reduction by decreasing the pump-line attenuation at the millikelvin stage of the refrigerator.

To achieve quantum-limited noise performance on an amplifier, the signal port input noise temperature $T^{\in,1}$ must be thermalized to the quantum noise level. 
The quantum limit of half a quantum of added noise for a phase-preserving parametric amplifier originates from the input quantum noise at the idler frequency ($\omega_i = \omega_p - \omega_s$). 
Having another port for pump power delivery, however, introduces additional channels for added noise. 

To quantify the effect of pump port input noise on amplifier noise performance, we inject noise along with the coherent pump tone into the pump line and measure the output noise of the amplifier, as illustrated in Fig.~\ref{fig:noise}(a). 
The injected noise is produced by mixing artificially generated noise — which has a flat-top power spectral density from DC to 1~GHz — with a local oscillator near signal and pump frequency, respectively. 
In this way, we separately investigate the effects of the noise in the signal and idler frequency bands (referred to as signal-band noise, the idler frequency for a degenerate amplifier is usually close to the signal frequency) and the noise in the pump frequency band (referred to as pump-band noise). 
The output noise power spectral density is measured at the frequency $\omega_s$ of 20~dB small-signal phase-preserving gain. 
Both the input and output noise power spectral densities are calibrated into noise temperatures $T_N[\omega]$ at the device ports (Appendix A). 
We perform this experiment on the F-SPA at $\omega_s/2\pi = 5.04$~GHz and on the C-SPA at $\omega_s/2\pi = 4.76$~GHz, corresponding to the operating points with highest $P_{\mathrm{1dB}}$ value for each device. 

Without noise injection, the output noise of both the F-SPA (marked with circles) and the C-SPA (marked with triangles) are close to the quantum limit, given by $T^{\out,1}_N[\omega_s] = G \cdot \hbar \omega_s / k_B$. 
Upon injecting noise into the signal and idler frequency bands, the output noise temperature increases linearly with the pump-port input noise temperature. 
As indicated by the yellow curved lines in Fig.~\ref{fig:noise}(b), the F-SPA is characterized by a fitted linear coefficient of 0.30, while the C-SPA exhibits a higher coefficient of 3.6.
This coefficient should equal to twice the signal transmission coefficient defined in Eq.\ref{eq:T} due to reciprocity of transmission at both the signal and idle frequencies. 
Compared to the C-SPA, the F-SPA exhibits an order of magnitude better suppression against signal-band noise from the pump line. 
Notably, the F-SPA maintains near quantum-limited performance even with noise injection up to 4~K, showing only a minimal increase in added noise amounting to 0.05 quanta. 
This robust performance indicates that the thermalization requirement for the pump line of F-SPA can be significantly relaxed compared to conventional designs.

\begin{figure}[htb]
\includegraphics[width = 8.6cm]{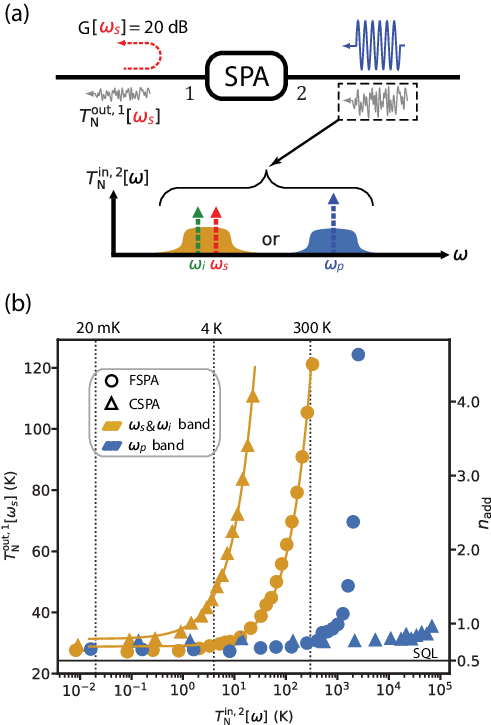}
\caption
{\label{fig:noise} 
Pump-line noise injection experiment. (a) Experimental setup: output noise temperature from the signal port is measured as a function of the pump-port input noise temperature, under the 20~dB gain pumping condition. (b) Data obtained from F-SPA (circles) and C-SPA (triangles) under injected noise in signal and idler frequency bands (yellow) and the pump frequency band (blue) respectively. Yellow lines represent linear fits to the data, with a fitted slope of 3.6 for the C-SPA and 0.30 for the F-SPA. The black horizontal line indicates the standard quantum limit (SQL) of the added noises. 
}
\end{figure}

When injecting noise at pump frequency, the output noise temperature of the F-SPA starts to increase drastically (i.e.\ beyond linear dependence) at a pump-band noise temperature of $10^3$~K, as shown in ~\ref{fig:noise}(b). 
In contrast, the output noise temperature of the C-SPA barely starts to increase at $10^5$~K of pump-band noise. 
This 20~dB difference in the threshold pump-band noise temperature is just another manifestation of the improved pump coupling efficiency on the F-SPA. 
However, the exact mechanism behind how pump-band noise causes signal-band output noise to increase requires further investigation.

Improving the pump coupling efficiency indeed increases the susceptibility of F-SPA to noise at the pump frequency. 
However, this should not degrade its noise performance in standard applications where the pump-band noise temperature typically remains well below 1000~K. 
Nonetheless, the filtering at the pump port of the F-SPA provides improved suppression of signal-band noise, thereby offering an opportunity for substantial heat load reduction. 
This reduction may be achieved through decreasing or potentially eliminating attenuation at the millikelvin stage of the dilution refrigerator. 

\end{section}
\begin{section}{Conclusion and discussion}

In this work, we have demonstrated improvements of 3 orders of magnitude in both power efficiency and pump-leakage suppression on a filter-coupled SNAIL parametric amplifier compared to its conventional capacitor-coupled counterpart. This improvement is achieved by implementing three-stage, on-chip filters on the signal and pump ports of the device. 
Most previous works incorporating microwave filters into JPA designs have targeted at improving the bandwidth and dynamic range~\cite{Broadband_JPA,Naaman2019,kaufman_Chebyshev2023,kaufman_SimpleHighSaturation2024}. 
Our work addresses the increasingly important issue of pump power delivery for three-wave-mixing JPAs. While a recent study ~\cite{kaufman_SimpleHighSaturation2024} discussed the use of an external diplexer to facilitate pump delivery, our work provides the methodology for optimizing the pump coupling at the device design level. The on-chip filtering strategy offers advantages in terms of compactness and eliminates insertion loss from additional external components. Moreover, the network-based description we introduced, particularly the formulation of the resonance condition in terms of Thevenin impedance, provides a systematic framework for designing the embedding circuit for Josephson elements with frequency-dependent coupling to the environment. We believe the strategy of including microwave filters in circuit QED design should benefit many other applications featuring off-resonant parametric drives~\cite{Kerr_Cat_2020,GKP_2020,zhouRealizingAlltoallCouplings2023,chapmanHighOnOffRatioBeamSplitterInteraction2023,xiaFastSuperconductingQubit2023}. 


Moreover, our work shows the benefits of the signal-pump frequency separation of three-wave mixing amplifiers, which allows for substituting pump line attenuation with filtering, reducing both the cryogenic power dissipation and the overall pump power required from the room-temperature rf generator. In particular, the filter-coupled SPA is robust against thermal noise on the pump port with up to 4~K noise temperature. This demonstration opens up the possibility of eliminating attenuation at the millikelvin stage of the dilution refrigerator, and introducing a separate port for pumping kinetic-inductance-based parametric amplifiers that can operate at higher temperatures~\cite{xuRadiativelyCooledQuantum2024}. Such optimizations will greatly alleviate the challenge of operating a large array of quantum-limited amplifiers in a large-scale superconducting quantum processor.

\end{section}

\begin{section}{Acknowledgement}
The authors would like to thank Jos\'e Aumentado for providing the shot-noise tunnel junction used for noise calibration. We acknowledge Nicholas Frattini and Benjamin Chapman for helpful discussions. W.D. thanks Luigi Frunzio, Robert Schoelkopf, and A. Douglas Stone for their advice on this project. This material is based upon work supported by the U.S. Department of Energy, Office of Science, National Quantum Information Science Research Centers, Co-design Center for Quantum Advantage (C2QA) under contract number DE-SC0012704. S.S. acknowledges support from the Air Force Office of Scientific Research (Grants No. FA9550-20-1-0177 and No. FA9550-22-1-0203) and the Army Research Office (Grants No. W911NF-23-1-0051, No. W911NF-23-1-0096, and No. W911NF-23-1-0251). The use of fabrication facilities was supported by the Yale Institute for Nanoscience and Quantum Engineering (YINQE) and the Yale SEAS Cleanroom. The views and conclusions contained in this document are those of the authors and should not be interpreted as representing the official policies, either expressed or implied, of the U.S. Government. The U.S. Government is authorized to reproduce and distribute reprints for Government purposes notwithstanding any copyright notation herein. 
\end{section}
\appendix
\begin{section}{Device fabrication and measurement setup}

The fabrication and packaging processes of our device are originally developed and introduced in Ref.~\cite{SPA,Kerr_free}. 
The aluminum strip (bright pattern in Fig.~\ref{fig:device}(b)) along with an array of $M=10$ SNAILs embedded inline is fabricated with a single-step e-beam lithography and liftoff process. 
Each SNAIL consists of three large Josephson junctions with critical current $I_c = 10.1$~\textmu A in parallel with one small junction with critical current $I_c = 0.8$~\textmu A, all formed by the Dolan bridge process, as shown in the SEM image in Fig.~\ref{fig:device}(b) inset. 

The microstrip ground plane is formed by 2~\textmu m thick layer of silver on the back of a 300~\textmu m thick silicon substrate. 
The silver back plane of the chip is glued using conducting silver paste to the copper back plane of a printed circuit board (PCB), which is soldered to a gold-plated aluminum box. The signal and pump aluminum transmission line traces on the chip are wire-bonded to the copper transmission line traces on the PCB, which are soldered to edge-mount SMA connectors.

\begin{figure*}[htb]
\includegraphics[width=17cm]{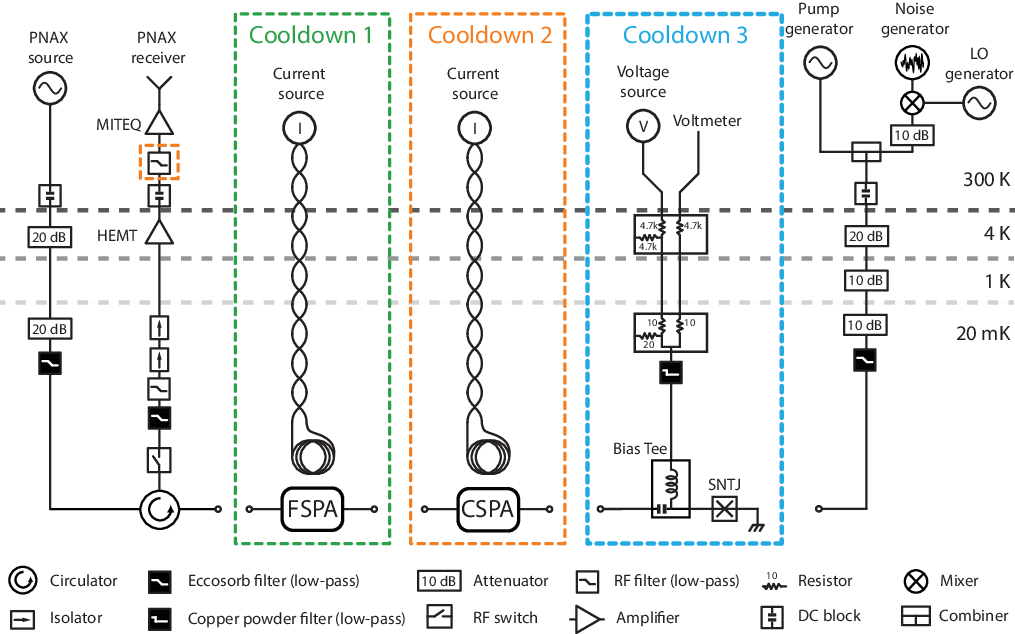}
\caption{\label{fig:wiring} 
Schematic of the cryogenic microwave wiring setup in this experiment. Parts of the setup enclosed in dashed boxes were connected in-line in separate cooldowns.}
\end{figure*}

A schematic of the cryogenic microwave measurement system used in this experiment is shown in \fig{fig:wiring}. All measurements were performed with a Keysight PNA-X N5242A network analyzer, with the scattering parameters measured with the SMC (Scaler Mixer/Converter) measurement class, and the noise temperature measured with the NF (Noise Figure) measurement class. 

The FSPA and CSPA were measured with the same wiring setup in two consecutive cooldowns, and in a third cooldown we used a Shot-Noise Tunnel Junction (SNTJ)~\cite{SNT} to calibrate the output chain. When measuring the CSPA device, a lowpass filter was added to the output chain in order to prevent the pump leakage from saturating the room temperature MITEQ amplifier. The two output line setups, with and without the lowpass filter, are calibrated respectively with the SNTJ. 

\begin{subsection}{Noise calibration}

\begin{figure}[htb]
\includegraphics[width=8.5cm]{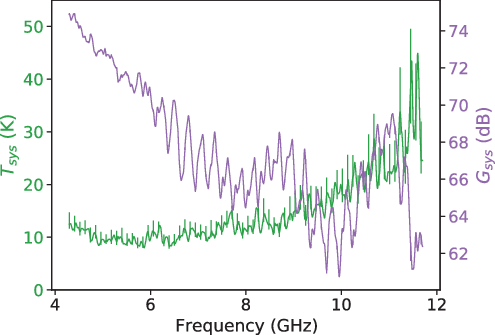}
\caption{\label{fig:SNT} 
Output line gain and noise temperature calibrated with a SNTJ. }
\end{figure}


The SNTJ serves as a self-calibrated noise source~\cite{SNT_Malnou2024} with which we can calibrate the gain $G_\sys$ and noise temperature $T_\sys$ of the output chain, shown in \fig{fig:SNT}. The uncertainty for noise temperature calibration is dominated by the impedance mismatch on the SNTJ and loss in the bias-tee. We accordingly extract the errorbar as a function of frequency using the same model as in \cite{JAMPA}. 

We then measured the noise visibility ratio (NVR) of the FSPA and CSPA, defined as the ratio between the noise power spectral density with the JPA on and off ($\mathrm{NVR} = P_N^{\mathrm{on}}/P_N^{\mathrm{off}}$. When measuring $P_N^{\mathrm{off}}$ the pump line noise injection is turned off along with the coherent pump). The JPA output noise $T_N^{\out,1}$ is extracted from: 
\begin{equation}
\mathrm{NVR} = \frac{G_\sys(T_\sys+T_N^{\out,1})}{G_\sys(T_\sys+T_Q)}
\end{equation}
using the $T_\sys$ calibrated by the SNTJ. 

The added noise temperature plotted in Fig.~\ref{fig:data} is obtained with the assumption that the signal port input noise is quantum limited: $T_N^{\out,1} = G(T_Q + T_\mathrm{add})$, which tends to be a pessimistic estimate for the JPA noise performance, since the input thermal noise gets combined with the amplifier added noise. 

\end{subsection}

\begin{subsection}{Pump noise injection}

In the noise injection experiment presented in Section V, we used a noise generater NOD-5200 to produce a stochastic rf signal with a flat spectrum up to 1~GHz, and used a Marki M8-0420HS mixer to up-convert its frequency. The center frequency of the noise spectrum (set by the LO frequency in the mixing process) is 600~MHz detuned from the signal or pump frequency, to minimize any unwanted effect from the LO leakage tone. A 10~dB attenuator is applied after the mixer to suppress the LO leakage as well. 

We measured the noise power spectrum before the combiner, and used the pump line attenuation measured at room temperature to calibrate the input noise temperature $T_N^{\in,2}$ at device level. 

\end{subsection}

\end{section}
\begin{section}{Scattering parameter and linear network analysis}

In this appendix we provide some details on linear network analysis that support the results we reported in Section II and III. 
In B.1 we derive the relations between wave components defined from the 3-port network modeled by Fig. \ref{fig:circuit}. In B.2 we show the simulation results of the signal-port filter and pump-port filter as in Fig. \ref{fig:device}, and discuss how they are related to the 3-port network parameters using the signal flow graph analysis.

\begin{subsection}{Scattering analysis for the 3-port network}

The scattering parameters for a network is defined as $S_{ij} = \left. \frac{V^{\out,i}}{V^{\in,j}} \right|_{V^{\in,k \ne j} = 0}$, the transfer function from port $j$ input to port $i$ output with all ports terminated by $Z_0$. 

For the 3-port network shown in Fig. \ref{figA:Thevenin}(a), port 3 is terminated by the Josephson dipole which we treat as a linear load. Therefore, the Josephson dipole reflects the port 3 outgoing wave back into the network: $V^{\in,3} = \Gamma_3 V^{\out,3}$.
So the total voltage across the two terminals of port 3 (i.e. across the Josephson dipole) is:
\begin{equation}
	V^{\J} = V^{\in,3} + V^{\out,3} = (1 + \Gamma_3) V^{\out,3}.
\end{equation}
Using the network scattering parameters, $V^{\out,3}$ can be evaluated as
\begin{equation}\label{eq:Vout3}
\begin{aligned}
	V^{\out,3} &= S_{33} V^{\in,3} + S_{31} V^{\in,1} + S_{32} V^{\in,2} \\
	&= \frac{ S_{31} }{1 - \Gamma_3 S_{33}} V^{\in,1} + \frac{ S_{32} }{1 - \Gamma_3 S_{33}}  V^{\in,2}.
\end{aligned}
\end{equation}

Since the pump is applied only from port 2 (i.e. $V^{\in,1}_{p} = 0$), the pump coupling efficiency from Eq.\eqref{eq:xipump} can be expressed in terms of scattering parameters as: 
\begin{equation}\label{eq:xipump1}
	\eta_p^{\mathrm{c}} = \abs{\frac{1 + \Gamma_3[\omega_p]}{1 - \Gamma_3[\omega_p] S_{33}[\omega_p]} S_{32}[\omega_p]}^2 \frac{Z_0}{\omega_p L_{\J}}.
\end{equation}

The signal, on the other hand, is applied only from port 1 (i.e. $V^{\in,2}_{s} = 0$). 
Signal power gain is expected in reflection from port 1, which is given by:
\begin{equation}\label{eq:G}
G = \abs{\frac{V^{\out,1}_{s} }{V^{\in,1}_{s} }}^2 = \abs{ S_{11}[\omega_s] + \frac{ S_{13}[\omega_s] \Gamma_3[\omega_s] S_{31}[\omega_s] }{1 - \Gamma_3[\omega_s] S_{33}[\omega_s]} }^2
\end{equation}
where $\Gamma_3[\omega_s]$ is the reflection coefficient from the Josephson dipole at the signal frequency. 
In our model, as illustrated in Fig.~\ref{fig:circuit}(c), signal amplification translates into a negative real part for the pumped admittance $Y^{\mathrm{nl}}(V^{\J}_{p})$. 
Furthermore, achieving high gain ($G \gg 1$) requires $1 - \Gamma_3[\omega_s] S_{33}[\omega_s] \approx 0$, which is satisfied for near-resonance signals. 

Evaluating $V^{\out,2}_s$ in a similar manner, we can also write the transmission coefficient from port 1 to port 2:
\begin{equation}\label{eq:T}
T = \abs{\frac{V^{\out,2}_{s} }{V^{\in,1}_{s} }}^2 = \abs{ S_{21}[\omega_s] +  \frac{ S_{23}[\omega_s] \Gamma_3[\omega_s] S_{31}[\omega_s] }{1 - \Gamma_3[\omega_s] S_{33}[\omega_s]} }^2.
\end{equation}
In order to achieve a noise performance approaching the quantum limit, the signal loss via port 2 should be minimized so that $T \ll G$, essentially demanding $\abs{S_{23}[\omega_s]} \ll \abs{S_{13}[\omega_s]}$. 

%

\end{subsection}

\begin{subsection}{Signal flow graph analysis for concatenating two-port networks}

In Section II we formulated the goals for circuit synthesis in terms of scattering parameters on a 3-port network. 
Since it is somewhat tricky to design the 3-port distributed-element circuit directly, we chose to decompose the circuit into separately designed signal-port and pump-port filter networks. The 2-port scattering matrix of the signal-port network is represented as
\begin{equation}
S^{(s)} = 
\left(
\begin{matrix}
r^{'(s)} & t^{(s)}  \\
t^{(s)} & r^{(s)}
\end{matrix}
\right)
\end{equation}
and similarly for the pump-port network, denoted as $S^{(p)}$, where $t$, $r$ and $r^{'}$ are labeled in the signal flow diagram in Fig. \ref{figA:Thevenin}(c). 

\begin{figure}[htb]
\includegraphics{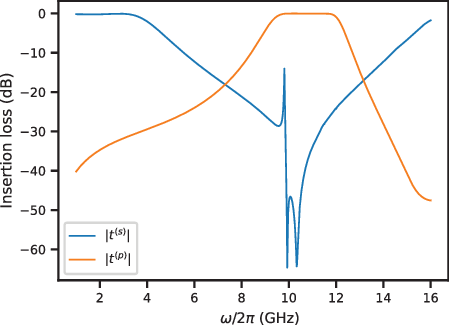}
\caption{\label{figA:Sparams} 
Simulated insertion loss of the signal-port filter and the pump-port filter respectively. }
\end{figure}

As discussed in Section III, the pump-port network is designed to pass the pump (around 10 GHz) and reject the signal (around 5 GHz), which is realized with a bandpass filter. The signal-port network is designed to create a stopband around pump frequency while placing the signal frequency within its transition band, for controlling the damping rate $\kappa_0$ of the resonant mode. 
The insertion loss of the two filter networks from AWR Axiem simulation are plotted in Fig.(\ref{figA:Sparams}). 

The standard filter designs typically aim to optimize the insertion loss $|t|$ and return loss $|r|$, assuming both ports are $Z_0$ matched. However, in our device the networks combine in the following manner: the Josephson dipole connects a pair of terminals, and the other pair of terminals are galvanically connected via the microstrip ground. 
Signal-flow graph analysis~\cite{Pozar} can be applied to Fig.~\ref{figA:Thevenin}(c) 
to construct the effective 3-port scattering matrix for the concatenated circuit.  

\begin{align}
	S_{31} &= \frac{2(1-r^{(p)})t^{(s)}}{3 - (r^{(s)} + r^{(p)}) - r^{(s)} r^{(p)}} \\
	S_{32} &= \frac{2(1-r^{(s)})t^{(p)}}{3 - (r^{(s)} + r^{(p)}) - r^{(s)} r^{(p)}} \\
        S_{21} &= \frac{2 t^{(s)} t^{(p)}}{3 - (r^{(s)} + r^{(p)}) - r^{(s)} r^{(p)}} \\
	S_{33} &= \frac{1 + (r^{(s)} + r^{(p)}) - 3 r^{(s)} r^{(p)}}{3 - (r^{(s)} + r^{(p)}) - r^{(s)} r^{(p)}}
\end{align}

In addition, the signal-flow graph analysis also gives us the linear response relations of interest between the voltage across the Josephson dipole and the input voltage, e.g.from the signal port, 
\begin{equation}
V^\J = \frac{\frac{2}{3}t^{(s)}(1 - r^{(p)})}{1 - \frac{1}{3} r^{(s)} r^{(p)} - \frac{1}{3}(r^{(s)} + r^{(p)})} V^{\in,1}.
\end{equation}

The most important consideration in our design process is the precise control of the resonant mode frequency and damping rate, which we will explain in detail in the next appendix. 

\begin{figure}[!ht]
\includegraphics{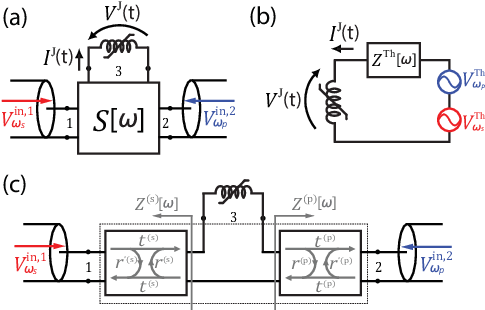}
\caption
{\label{figA:Thevenin}
Three equivalent descriptions of a Josephson-dipole-based parametric amplifier, modeling the linear circuit by (a) a 3-port network (b) a Thevenin equivalent circuit (c) the combination of 2-port networks. 
}
\end{figure}

\end{subsection}

\end{section}
\begin{section}{Resonance condition from Thevenin circuit method}

In this appendix, we formulate the resonance condition for a general circuit containing a lumped Josephson dipole, such as the example shown in Fig.~\ref{figA:Thevenin}(a). 
Our method can efficiently extract the resonance frequency, damping rate, and \emph{Josephson participation ratio}~\cite{minevEnergyparticipationQuantizationJosephson2021} of a classical mode, whose Kerr nonlinearity is much smaller than its damping rate. We expect our method to provide some insights into designing Josephson-element-based parametric amplifiers and couplers with complicated embedding circuits. 

According to the AC Thevenin theorem, the two-terminal linear bilateral circuit external to the Josephson dipole can be replaced by a Thevenin equivalent circuit consisting of a single impedance $Z^{\mathrm{Th}}[\omega]$ in series with a single voltage source $V^{\mathrm{Th}}(t)$, as depicted in Fig.~\ref{figA:Thevenin}(b). The external voltage source has only two harmonic components (i.e. signal and pump) and thus $V^{\mathrm{Th}}(t) = \Re (V^{\mathrm{Th}}_{\omega_s} e^{j \omega_s t} + V^{\mathrm{Th}}_{\omega_p} e^{j \omega_p t} ) $. 

For obtaining the linear eigenmode of the system, we replace the nonlinear Josephson dipole by its inductance $L_{\J}$, with a linear response: $V^{\J} = L_{\J} \Dif{I^{\J}}{t}$. To find the `undriven' modes of motion allowed by a homogeneous linear system, a Laplace transform (represented by $\mathcal{L}$) can be applied to the equation of motion~\cite{oppenheim_ss}: 
\begin{equation}
s L_{\J} \mathcal{L} \{ I^{\J} (t)\} + Z^{\mathrm{Th}}(s) \mathcal{L} \{ I^{\J} (t)\} = 0
\end{equation}
This gives rise to the \emph{resonance condition}, shown as \eq{eq:resonance} in Section II. The general solution to the system can always be decomposed into a linear combination of eigenmode excitations: 
\begin{equation}
	I^{\J}(t) = \sum_a I_{a} \exp{- \frac{\kappa_a}{2} t + j \omega_a t}
\end{equation}
using the discrete set of solutions $\{ s_a = j \omega_a - \frac{\kappa_a}{2} \}$ to \eq{eq:resonance}. 

Next, we bridge between Laplace domain functions and Fourier domain functions, in order to formulate the resonance condition in terms of Fourier domain response function $Z^\mathrm{Th}[\omega]$ which is easier to obtain from linear circuit simulation. 
We use the notation that the Fourier domain function $f[\omega]$ equals the Laplace domain function $f(s)$ taken on the imaginary axis $s = j \omega$. 
Note that we have the following identity with respect to their derivatives: 
\begin{equation}\label{eq:fprime}
\begin{aligned}
f'[\omega] &:= \frac{f[\omega + \mathrm{d} \omega] - f[\omega]}{\mathrm{d} \omega} \\
&= \frac{f(j\omega + j\mathrm{d} \omega) - f(j\omega)}{\mathrm{d} \omega} \\
&=: j f'(j\omega)
\end{aligned}
\end{equation}

Assuming $\kappa_a \ll \omega_a$, the following approximation from Taylor expansion: 
$$
f(j\omega_a - \frac{\kappa_a}{2}) \approx f(j\omega_a) - \frac{\kappa_a}{2} f'(j\omega)
$$
holds for any analytic function $f(s)$. Therefore, we can write: 
\begin{equation}
	Z^{\mathrm{Th}}(j \omega_a - \frac{\kappa_a}{2} ) \approx Z^{\mathrm{Th}}[\omega_a] + j \frac{\kappa_a}{2} Z^{\mathrm{Th}'}[\omega_a] 
\end{equation}

Plugging in the resonance condition Eq. \ref{eq:resonance} in the main text, i.e. $Z^{\mathrm{Th}}(j \omega_a - \frac{\kappa_a}{2} ) = - j \omega_a L_{\J} + \frac{\kappa_a}{2} L_{\J} $: 

\begin{subequations}\label{eq:Z0}
\begin{align}
\Im Z^{\mathrm{Th}}[\omega_a] &= - \left(\omega_a L_{\J} + \frac{\kappa_a}{2} \Re Z^{\mathrm{Th}'}[\omega_a] \right)  \label{eq:Z1}\\
\Re Z^{\mathrm{Th}}[\omega_a] &= \frac{\kappa_a}{2} \left( L_{\J} + \Im Z^{\mathrm{Th}'}[\omega_a] \right) \label{eq:Z2}
\end{align}
\end{subequations}
These are the requirements on $Z^{\mathrm{Th}}[\omega]$ (both real and imaginary part) that we synthesize for a resonance mode with frequency $\omega_a/2\pi$ and linewidth $\kappa_a/2\pi$. 

The Thevenin impedance for our design can either be expressed in terms of the 3-port network scattering parameter:
\begin{equation}
	Z^\mathrm{Th}[\omega] = Z_0 \frac{1 + S_{33}[\omega] }{1 - S_{33}[\omega]}
\end{equation}
or be separated into two parts: $Z^\mathrm{Th}[\omega] = Z^{(s)}[\omega] + Z^{(p)}[\omega]$ where
\begin{subequations}
\begin{align}
	Z^{(s)}[\omega] &= Z_0 \frac{1 + r^{(s)}[\omega] }{1 - r^{(s)}[\omega]} \\
	Z^{(p)}[\omega] &= Z_0 \frac{1 + r^{(p)}[\omega] }{1 - r^{(p)}[\omega]}
\end{align}
\end{subequations}
are the output impedance towards signal and pump port respectively, as indicated in Fig. \ref{figA:Thevenin}(c). 

\begin{figure}[htb]
\includegraphics{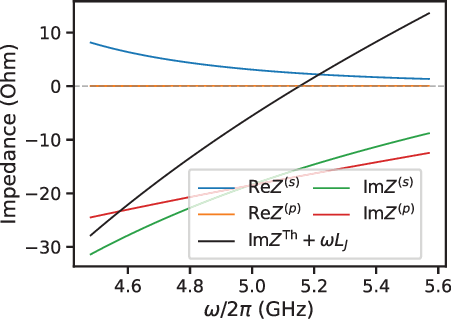}
\caption{
\label{figA:Zout} 
Simulated output impedance towards signal port $Z^{(s)}$ and output impedance towards pump port $Z^{(p)}$. The resonance frequency is determined by the point where the black curve (an example with $L_{\J}(\Phi_{\mathrm{ext}})$ evaluated at $\Phi_{\mathrm{ext}} = 0.3 \Phi_{0}$) crosses 0. The slope at this point is related to the Josephson participation of the mode. Real part of $Z^{(s)}$ evaluated at the resonance frequency determines the linewidth of the mode. }
\end{figure}

In Fig.~\ref{figA:Zout}, we show the $Z^{(s)}$ and $Z^{(p)}$ simulated (using AWR Axiem) around the desired resonance frequency. 
We also plot the imaginary part of $Z^{(s)}[\omega] + Z^{(p)}[\omega] + j \omega L_{\J}(\Phi_{\mathrm{ext}})$ evaluated at $\Phi_{\mathrm{ext}} = 0.3 \Phi_{0}$ as the black curve. 
As the second term on the right hand side of Eq.(\ref{eq:Z1}) is negligible in our case, the resonance frequency $\omega_a$ is essentially the point where the black curve crosses 0. 
We then evaluate $\kappa_a$ from Eq.(\ref{eq:Z2}), using $\Re Z^{\mathrm{Th}}$ and $\Im Z^{\mathrm{Th}'}$ taken at $\omega_a$. 
Since the real part of $Z^{(p)}$ is almost 0 around resonance frequency (as the pump-port filter rejects the on-resonance signal), the damping rate of the mode originates entirely from the dissipation at the signal port. 

In Fig.~\ref{fig:device}(c) from the main text we have shown the resonance frequency and linewidth extracted using the method above, with varying $L_{\J} (\Phi_{\mathrm{ext}})$ for the SNAIL array\cite{SNAIL}. They agree with the experimental results fairly well. 

Lastly, we can also extract the \emph{Josephson inductive participation ratio} of each mode $a$ from the Thevenin circuit method. Following the same ideas as blackbox circuit quantization~\cite{niggBlackBoxSuperconductingCircuit2012}, the Josephson dipole can be expressed as a linear inductor $L_{\J}$ in parallel with a purely nonlinear element. Under the assumption of weak dissipation (i.e. $\kappa_a \ll \omega_a$) for all eigenmodes of interest, the full linear circuit consisting of $L_{\J}$ in parallel with the Thevenin impedance can be decomposed into a set of poles in the form of parallel RLCs, i.e. an approximate Foster-equivalent admittance.  
\begin{equation}
\frac{1}{s L_{\J}} + \frac{1}{Z^{\mathrm{Th}}(s)} = \sum_a \left(s C_a + \frac{1}{s L_a} + \frac{1}{R_a} \right)
\end{equation}
Combining this with the resonance condition Eq. \ref{eq:resonance}, we can arrive at the following result: 
\begin{equation}\label{eq:pJ}
p_a^\J := \frac{L_a}{L_{\J}} \approx \frac{2}{1 + \frac{\Im Z^{\mathrm{Th}'}[\omega_a]}{L_{\J}}}
\end{equation}
which by definition is the inductive participation ratio (twice the energy participation ratio as defined in~\cite{minevEnergyparticipationQuantizationJosephson2021}) of the Josephson dipole in mode $a$. 

An observation is made that $p_a^\J$ enters \eq{eq:Z2} such that $\Re Z^{\mathrm{Th}}[\omega_a] / L_{\J} = \kappa_a / p_a^\J$. It can be understood as a scaling factor between the damping rate at the Josephson dipole and that for the the resonance mode. 

\end{section}
\begin{section}{Modeling the pumped Josephson dipole}

In this appendix, we introduce a circuit model describing a current-pumped parametric process in a single lumped Josephson dipole, in analogy with previous work on flux-pumped parametric processes~\cite{Sundqvist2014}. 
This description applies to the synthesis of parametric amplifiers and converters, as well as of the current-pumped ancilla for gaussian operations on bosonic modes~\cite{gaoEntanglementBosonicModes2019,chapmanHighOnOffRatioBeamSplitterInteraction2023,dingQuantumControlOscillator2024}. 

Similar to the Taylor expansion of the Josephson relation $I = \frac{\phi_0}{L_{\J}} \sin \varphi$ for a single Josephson junction, the electric current $I^{\J}(t)$ is linked to the superconducting phase $\varphi^{\J}(t) = \Phi^{\J}(t)/\phi_0$ across an arbitrary Josephson dipoles by the following relation:
\begin{equation}\label{eq:Iphi}
	I^{\J} = \frac{\phi_0}{L_{\J}}\( \varphi^{\J} +  \frac{c_3}{2} \varphi^{\J2}  + \cdots \)
\end{equation}
where $L_{\J}$ is the total inductance of the Josephson dipole. Derivation of $L_{\J}$ and the expansion coefficient $c_3$ depends on the physical realization of the dipole, examples are present for a lumped array of SNAIL~\cite{SNAIL}, rf SQUID or Gradiometric-SNAIL~\cite{G-SNAIL}. 
Here, we have kept only the 3rd order nonlinearity and study the parametric squeezing processes arising from this term. 

Consider a pump at $\omega_p = \omega_s + \omega_i$ applied to activate squeezing between two temporal modes centered at $\omega_s$ and $\omega_i$ respectively (which holds for phase-preserving amplification). We decompose $I^{\J}(t)$ and $\varphi^{\J}(t)$ in \eq{eq:Iphi} using the phasor representation: 
\begin{equation}
\begin{aligned}
	I^{\J}(t) &= \Re \( I^{\J}_{p} \exp{j\omega_p t} + I^{\J}_{s} \exp{j\omega_s t} + I^{\J}_{i} \exp{j\omega_i t} \) \\
	\varphi^{\J}(t) &= \Re \( \varphi^{\J}_{p} \exp{j\omega_p t} + \varphi^{\J}_{s} \exp{j\omega_s t} + \varphi^{\J}_{i} \exp{j\omega_i t} \)
\end{aligned}
\end{equation}
Plugging it into \eq{eq:Iphi}, the following relations are obtained from harmonic balance: 
\begin{equation}\label{2}
\begin{aligned}
	I^{\J}_{s} &= \frac{\phi_0}{L_{\J}} \(\varphi^{\J}_{s} + \frac{c_3}{2} \varphi^{\J}_{p} \varphi^{\J*}_{i}\) \\
	I^{\J*}_{i} &= \frac{\phi_0}{L_{\J}} \(\varphi^{\J*}_{i} + \frac{c_3}{2} \varphi^{\J*}_{p} \varphi^{\J}_{s}\) 
\end{aligned}
\end{equation}

The stiff-pump approximation assumes that the existence of probes at other frequencies does not affect the steady state $V^{\J}_{\omega_p}$ evaluated from linear circuit analysis. Under such an approximation, the harmonic component $\varphi^{\J}_{p} = \frac{V^{\J}_{p} }{j \omega_p \phi_0}$ can be treated as a \emph{parameter} (instead of a \emph{variable}). The harmonic balance \eq{2} between $\omega_s$ and $\omega_i$ harmonic components is thus linearizable, which can be written as a cross-admittance matrix: 
\begin{equation}\label{eq:cross_admittance}
\(
\begin{matrix}
I^{\J}_{s}\\
I^{\J*}_{i}
\end{matrix}
\)
= 
\frac{1}{j L_{\J}}
\(
\begin{matrix}
\frac{1}{\omega_s} & \frac{ \frac{c_3}{2}\varphi^{\J}_{p} }{-\omega_i}  \\
\frac{\frac{c_3}{2}\varphi^{\J*}_{p}}{\omega_s} & \frac{1}{-\omega_i}
\end{matrix}
\)
\(
\begin{matrix}
V^{\J}_{s}\\
V^{\J*}_{i}
\end{matrix}
\)
\end{equation}
where the off-diagonal term arises from the pump and takes the steady state $\varphi^{\J}_{p}$ as a parameter. 

In the case of phase-preserving parametric amplification, there is an external signal probe at $\omega_s$ only, while no probe is applied at the idler frequency $\omega_i$, as indicated in Fig. (\ref{figA:Thevenin}). Therefore, the $\omega_i$ harmonic component at the Josephson dipole has to satisfy a constraint posed by the passive response of the linear coupling circuit, which can be formulated as: 
\begin{equation}\label{eq:idler}
	V^{\J}_{i} + Z^{\mathrm{Th}}[\omega_i] I^{\J}_{i} = 0
\end{equation}

Solving equation (\ref{eq:cross_admittance}) given this constraint, we arrive at a self-consistent same-frequency response to the signal at the pumped Josephson dipole: 
\begin{equation}\label{eq:s1}
\left. \frac{I^{\J}_{s}}{V^{\J}_{s}} \right|_{ \rm{(\ref{eq:idler})}} = \frac{1}{j \omega_s L_{\J}} + \frac{1}{j L_\J \omega_s } \frac{ \abs{ \frac{c_3}{2} \varphi^{\J}_{p} }^2 }{ \frac{j L_\J \omega_i}{Z^{\mathrm{Th}*}[\omega_i]} -1}
\end{equation}
where the second term is labeled $Y^{\mathrm{nl}}\(V^{\J}_{p}\)$ in the main text, an effective admittance that arises upon pumping as illustrated in Figure 2(c). 

Lastly, as we commented in Section II, we need a negative real part from $Y^{\mathrm{nl}}\left(V^{\J}_{p}\right)$
in order for the device to serve as an amplifier. 
That is achieved if the idler ($\omega_i$) is on-resonance with a standing mode, i.e. when $Z^{\mathrm{Th}}[\omega_i]$ satisfies equations (\ref{eq:Z0}) as explained in the previous appendix. 
Using the definition in \eq{eq:pJ}, the condition for the idler being on-resonance can be rewritten as
\begin{equation}
Z^{\mathrm{Th}}[\omega_i] \approx - j \omega_i L_{\J} \( 1 + j \frac{1}{p^{\J}_i Q_i} \)
\end{equation}
where $Q_i$ is quality factor of the standing mode that the idler tone resides in, and $p^{\J}_i$ is Josephson participation of this mode. 

As a result, 
\begin{equation}\label{eq:negativeY}
\begin{aligned}
    Y^{\mathrm{nl}}\(V^{\J}_{p}\) &= \frac{1}{j \omega_s L_{\J}} \frac{\abs{\frac{c_3}{2}\varphi^{\J}_{p}}^2}{ \frac{1}{1 - j \frac{1}{p^\J_i Q_i} } - 1 } \\
    & \approx - \frac{p^\J_i Q_i}{\omega_s L_{\J}} \abs{\frac{c_3}{2}\varphi^{\J}_{p}}^2
\end{aligned}
\end{equation}
which presents a \emph{negative conductance} viewed at the signal frequency. The last approximation holds under the assumption $p^{\J}_i Q_i \gg 1$, which is a well acknowledged claim that the $pQ$-product need to be greater than 1 for the amplifier to operate. 

As an example, our device is a degenerate amplifier where the idler tone and the signal tone populates the same resonance mode. Under the flux point $\Phi_{\mathrm{ext}} = 0.3 \Phi_{0}$ (the black curve in Fig.~\ref{figA:Zout}), the mode has a resonance frequency $\omega_a/2\pi = 5.15$ GHz and linewidth $\kappa_a/2\pi = 137$ MHz, with Josephson participation ratio $p^{\J}_a = 0.36\%$. This corresponds to $p^{\J}_a Q_a = 13.5$. 

\end{section}
%


\begin{thebibliography}{47}%
\makeatletter
\providecommand \@ifxundefined [1]{%
 \@ifx{#1\undefined}
}%
\providecommand \@ifnum [1]{%
 \ifnum #1\expandafter \@firstoftwo
 \else \expandafter \@secondoftwo
 \fi
}%
\providecommand \@ifx [1]{%
 \ifx #1\expandafter \@firstoftwo
 \else \expandafter \@secondoftwo
 \fi
}%
\providecommand \natexlab [1]{#1}%
\providecommand \enquote  [1]{``#1''}%
\providecommand \bibnamefont  [1]{#1}%
\providecommand \bibfnamefont [1]{#1}%
\providecommand \citenamefont [1]{#1}%
\providecommand \href@noop [0]{\@secondoftwo}%
\providecommand \href [0]{\begingroup \@sanitize@url \@href}%
\providecommand \@href[1]{\@@startlink{#1}\@@href}%
\providecommand \@@href[1]{\endgroup#1\@@endlink}%
\providecommand \@sanitize@url [0]{\catcode `\\12\catcode `\$12\catcode
  `\&12\catcode `\#12\catcode `\^12\catcode `\_12\catcode `\%12\relax}%
\providecommand \@@startlink[1]{}%
\providecommand \@@endlink[0]{}%
\providecommand \url  [0]{\begingroup\@sanitize@url \@url }%
\providecommand \@url [1]{\endgroup\@href {#1}{\urlprefix }}%
\providecommand \urlprefix  [0]{URL }%
\providecommand \Eprint [0]{\href }%
\providecommand \doibase [0]{http://dx.doi.org/}%
\providecommand \selectlanguage [0]{\@gobble}%
\providecommand \bibinfo  [0]{\@secondoftwo}%
\providecommand \bibfield  [0]{\@secondoftwo}%
\providecommand \translation [1]{[#1]}%
\providecommand \BibitemOpen [0]{}%
\providecommand \bibitemStop [0]{}%
\providecommand \bibitemNoStop [0]{.\EOS\space}%
\providecommand \EOS [0]{\spacefactor3000\relax}%
\providecommand \BibitemShut  [1]{\csname bibitem#1\endcsname}%
\let\auto@bib@innerbib\@empty
\bibitem [{\citenamefont {Roy}\ and\ \citenamefont
  {Devoret}(2018)}]{JPA_review}%
  \BibitemOpen
  \bibfield  {author} {\bibinfo {author} {\bibfnamefont {Ananda}\ \bibnamefont
  {Roy}}\ and\ \bibinfo {author} {\bibfnamefont {Michel}\ \bibnamefont
  {Devoret}},\ }\bibfield  {title} {\enquote {\bibinfo {title} {Quantum-limited
  parametric amplification with josephson circuits in the regime of pump
  depletion},}\ }\href {\doibase 10.1103/PhysRevB.98.045405} {\bibfield
  {journal} {\bibinfo  {journal} {Phys. Rev. B}\ }\textbf {\bibinfo {volume}
  {98}},\ \bibinfo {pages} {045405} (\bibinfo {year} {2018})}\BibitemShut
  {NoStop}%
\bibitem [{\citenamefont {Jeffrey}\ \emph {et~al.}(2014)\citenamefont
  {Jeffrey}, \citenamefont {Sank}, \citenamefont {Mutus}, \citenamefont
  {White}, \citenamefont {Kelly}, \citenamefont {Barends}, \citenamefont
  {Chen}, \citenamefont {Chen}, \citenamefont {Chiaro}, \citenamefont
  {Dunsworth}, \citenamefont {Megrant}, \citenamefont {O'Malley}, \citenamefont
  {Neill}, \citenamefont {Roushan}, \citenamefont {Vainsencher}, \citenamefont
  {Wenner}, \citenamefont {Cleland},\ and\ \citenamefont
  {Martinis}}]{Jeffrey_2014_FastReadout}%
  \BibitemOpen
  \bibfield  {author} {\bibinfo {author} {\bibfnamefont {Evan}\ \bibnamefont
  {Jeffrey}}, \bibinfo {author} {\bibfnamefont {Daniel}\ \bibnamefont {Sank}},
  \bibinfo {author} {\bibfnamefont {J.~Y.}\ \bibnamefont {Mutus}}, \bibinfo
  {author} {\bibfnamefont {T.~C.}\ \bibnamefont {White}}, \bibinfo {author}
  {\bibfnamefont {J.}~\bibnamefont {Kelly}}, \bibinfo {author} {\bibfnamefont
  {R.}~\bibnamefont {Barends}}, \bibinfo {author} {\bibfnamefont
  {Y.}~\bibnamefont {Chen}}, \bibinfo {author} {\bibfnamefont {Z.}~\bibnamefont
  {Chen}}, \bibinfo {author} {\bibfnamefont {B.}~\bibnamefont {Chiaro}},
  \bibinfo {author} {\bibfnamefont {A.}~\bibnamefont {Dunsworth}}, \bibinfo
  {author} {\bibfnamefont {A.}~\bibnamefont {Megrant}}, \bibinfo {author}
  {\bibfnamefont {P.~J.~J.}\ \bibnamefont {O'Malley}}, \bibinfo {author}
  {\bibfnamefont {C.}~\bibnamefont {Neill}}, \bibinfo {author} {\bibfnamefont
  {P.}~\bibnamefont {Roushan}}, \bibinfo {author} {\bibfnamefont
  {A.}~\bibnamefont {Vainsencher}}, \bibinfo {author} {\bibfnamefont
  {J.}~\bibnamefont {Wenner}}, \bibinfo {author} {\bibfnamefont {A.~N.}\
  \bibnamefont {Cleland}}, \ and\ \bibinfo {author} {\bibfnamefont {John~M.}\
  \bibnamefont {Martinis}},\ }\bibfield  {title} {\enquote {\bibinfo {title}
  {Fast accurate state measurement with superconducting qubits},}\ }\href
  {\doibase 10.1103/PhysRevLett.112.190504} {\bibfield  {journal} {\bibinfo
  {journal} {Phys. Rev. Lett.}\ }\textbf {\bibinfo {volume} {112}},\ \bibinfo
  {pages} {190504} (\bibinfo {year} {2014})}\BibitemShut {NoStop}%
\bibitem [{\citenamefont {Bienfait}\ \emph {et~al.}(2016)\citenamefont
  {Bienfait}, \citenamefont {Pla}, \citenamefont {Kubo}, \citenamefont {Stern},
  \citenamefont {Zhou}, \citenamefont {Lo}, \citenamefont {Weis}, \citenamefont
  {Schenkel}, \citenamefont {Thewalt}, \citenamefont {Vion}, \citenamefont
  {Esteve}, \citenamefont {Julsgaard}, \citenamefont {Mølmer}, \citenamefont
  {Morton},\ and\ \citenamefont {Bertet}}]{Bienfait_2016_ESR}%
  \BibitemOpen
  \bibfield  {author} {\bibinfo {author} {\bibfnamefont {A.}~\bibnamefont
  {Bienfait}}, \bibinfo {author} {\bibfnamefont {J.~J.}\ \bibnamefont {Pla}},
  \bibinfo {author} {\bibfnamefont {Y.}~\bibnamefont {Kubo}}, \bibinfo {author}
  {\bibfnamefont {M.}~\bibnamefont {Stern}}, \bibinfo {author} {\bibfnamefont
  {X.}~\bibnamefont {Zhou}}, \bibinfo {author} {\bibfnamefont {C.~C.}\
  \bibnamefont {Lo}}, \bibinfo {author} {\bibfnamefont {C.~D.}\ \bibnamefont
  {Weis}}, \bibinfo {author} {\bibfnamefont {T.}~\bibnamefont {Schenkel}},
  \bibinfo {author} {\bibfnamefont {M.~L.W.}\ \bibnamefont {Thewalt}}, \bibinfo
  {author} {\bibfnamefont {D.}~\bibnamefont {Vion}}, \bibinfo {author}
  {\bibfnamefont {D.}~\bibnamefont {Esteve}}, \bibinfo {author} {\bibfnamefont
  {B.}~\bibnamefont {Julsgaard}}, \bibinfo {author} {\bibfnamefont
  {K.}~\bibnamefont {Mølmer}}, \bibinfo {author} {\bibfnamefont {J.~J.L.}\
  \bibnamefont {Morton}}, \ and\ \bibinfo {author} {\bibfnamefont
  {P.}~\bibnamefont {Bertet}},\ }\bibfield  {title} {\enquote {\bibinfo {title}
  {Reaching the quantum limit of sensitivity in electron spin resonance},}\
  }\href {\doibase 10.1038/nnano.2015.282} {\bibfield  {journal} {\bibinfo
  {journal} {Nature Nanotechnology}\ } (\bibinfo {year} {2016}),\
  10.1038/nnano.2015.282}\BibitemShut {NoStop}%
\bibitem [{\citenamefont {Brubaker}\ \emph {et~al.}(2017)\citenamefont
  {Brubaker}, \citenamefont {Zhong}, \citenamefont {Gurevich}, \citenamefont
  {Cahn}, \citenamefont {Lamoreaux}, \citenamefont {Simanovskaia},
  \citenamefont {Root}, \citenamefont {Lewis}, \citenamefont {Al~Kenany},
  \citenamefont {Backes}, \citenamefont {Urdinaran}, \citenamefont {Rapidis},
  \citenamefont {Shokair}, \citenamefont {van Bibber}, \citenamefont {Palken},
  \citenamefont {Malnou}, \citenamefont {Kindel}, \citenamefont {Anil},
  \citenamefont {Lehnert},\ and\ \citenamefont {Carosi}}]{Brubaker_2017_Axion}%
  \BibitemOpen
  \bibfield  {author} {\bibinfo {author} {\bibfnamefont {B.~M.}\ \bibnamefont
  {Brubaker}}, \bibinfo {author} {\bibfnamefont {L.}~\bibnamefont {Zhong}},
  \bibinfo {author} {\bibfnamefont {Y.~V.}\ \bibnamefont {Gurevich}}, \bibinfo
  {author} {\bibfnamefont {S.~B.}\ \bibnamefont {Cahn}}, \bibinfo {author}
  {\bibfnamefont {S.~K.}\ \bibnamefont {Lamoreaux}}, \bibinfo {author}
  {\bibfnamefont {M.}~\bibnamefont {Simanovskaia}}, \bibinfo {author}
  {\bibfnamefont {J.~R.}\ \bibnamefont {Root}}, \bibinfo {author}
  {\bibfnamefont {S.~M.}\ \bibnamefont {Lewis}}, \bibinfo {author}
  {\bibfnamefont {S.}~\bibnamefont {Al~Kenany}}, \bibinfo {author}
  {\bibfnamefont {K.~M.}\ \bibnamefont {Backes}}, \bibinfo {author}
  {\bibfnamefont {I.}~\bibnamefont {Urdinaran}}, \bibinfo {author}
  {\bibfnamefont {N.~M.}\ \bibnamefont {Rapidis}}, \bibinfo {author}
  {\bibfnamefont {T.~M.}\ \bibnamefont {Shokair}}, \bibinfo {author}
  {\bibfnamefont {K.~A.}\ \bibnamefont {van Bibber}}, \bibinfo {author}
  {\bibfnamefont {D.~A.}\ \bibnamefont {Palken}}, \bibinfo {author}
  {\bibfnamefont {M.}~\bibnamefont {Malnou}}, \bibinfo {author} {\bibfnamefont
  {W.~F.}\ \bibnamefont {Kindel}}, \bibinfo {author} {\bibfnamefont {M.~A.}\
  \bibnamefont {Anil}}, \bibinfo {author} {\bibfnamefont {K.~W.}\ \bibnamefont
  {Lehnert}}, \ and\ \bibinfo {author} {\bibfnamefont {G.}~\bibnamefont
  {Carosi}},\ }\bibfield  {title} {\enquote {\bibinfo {title} {First results
  from a microwave cavity axion search at $24\text{ }\text{
  }\ensuremath{\mu}\mathrm{eV}$},}\ }\href {\doibase
  10.1103/PhysRevLett.118.061302} {\bibfield  {journal} {\bibinfo  {journal}
  {Phys. Rev. Lett.}\ }\textbf {\bibinfo {volume} {118}},\ \bibinfo {pages}
  {061302} (\bibinfo {year} {2017})}\BibitemShut {NoStop}%
\bibitem [{\citenamefont {Roy}\ \emph {et~al.}(2015)\citenamefont {Roy},
  \citenamefont {Kundu}, \citenamefont {Chand}, \citenamefont {Vadiraj},
  \citenamefont {Ranadive}, \citenamefont {Nehra}, \citenamefont {Patankar},
  \citenamefont {Aumentado}, \citenamefont {Clerk},\ and\ \citenamefont
  {Vijay}}]{Broadband_JPA}%
  \BibitemOpen
  \bibfield  {author} {\bibinfo {author} {\bibfnamefont {Tanay}\ \bibnamefont
  {Roy}}, \bibinfo {author} {\bibfnamefont {Suman}\ \bibnamefont {Kundu}},
  \bibinfo {author} {\bibfnamefont {Madhavi}\ \bibnamefont {Chand}}, \bibinfo
  {author} {\bibfnamefont {A~M}\ \bibnamefont {Vadiraj}}, \bibinfo {author}
  {\bibfnamefont {A}~\bibnamefont {Ranadive}}, \bibinfo {author} {\bibfnamefont
  {N}~\bibnamefont {Nehra}}, \bibinfo {author} {\bibfnamefont {Meghan~P}\
  \bibnamefont {Patankar}}, \bibinfo {author} {\bibfnamefont {J}~\bibnamefont
  {Aumentado}}, \bibinfo {author} {\bibfnamefont {A~A}\ \bibnamefont {Clerk}},
  \ and\ \bibinfo {author} {\bibfnamefont {R}~\bibnamefont {Vijay}},\
  }\bibfield  {title} {\enquote {\bibinfo {title} {{Broadband parametric
  amplification with impedance engineering: Beyond the gain-bandwidth
  product}},}\ }\href {\doibase 10.1063/1.4939148} {\bibfield  {journal}
  {\bibinfo  {journal} {Applied Physics Letters}\ }\textbf {\bibinfo {volume}
  {107}},\ \bibinfo {pages} {262601} (\bibinfo {year} {2015})}\BibitemShut
  {NoStop}%
\bibitem [{\citenamefont {Eichler}\ and\ \citenamefont
  {Wallraff}(2014)}]{Eichler2014}%
  \BibitemOpen
  \bibfield  {author} {\bibinfo {author} {\bibfnamefont {Christopher}\
  \bibnamefont {Eichler}}\ and\ \bibinfo {author} {\bibfnamefont {Andreas}\
  \bibnamefont {Wallraff}},\ }\bibfield  {title} {\enquote {\bibinfo {title}
  {{Controlling the dynamic range of a josephson parametric amplifier}},}\
  }\href {\doibase 10.1140/epjqt2} {\bibfield  {journal} {\bibinfo  {journal}
  {EPJ Quantum Technology}\ }\textbf {\bibinfo {volume} {1}},\ \bibinfo {pages}
  {1--19} (\bibinfo {year} {2014})}\BibitemShut {NoStop}%
\bibitem [{\citenamefont {Liu}\ \emph {et~al.}(2017)\citenamefont {Liu},
  \citenamefont {Chien}, \citenamefont {Cao}, \citenamefont {Lanes},
  \citenamefont {Alpern}, \citenamefont {Pekker},\ and\ \citenamefont
  {Hatridge}}]{Saturation_Gang}%
  \BibitemOpen
  \bibfield  {author} {\bibinfo {author} {\bibfnamefont {G.}~\bibnamefont
  {Liu}}, \bibinfo {author} {\bibfnamefont {T.-C.}\ \bibnamefont {Chien}},
  \bibinfo {author} {\bibfnamefont {X.}~\bibnamefont {Cao}}, \bibinfo {author}
  {\bibfnamefont {O.}~\bibnamefont {Lanes}}, \bibinfo {author} {\bibfnamefont
  {E.}~\bibnamefont {Alpern}}, \bibinfo {author} {\bibfnamefont
  {D.}~\bibnamefont {Pekker}}, \ and\ \bibinfo {author} {\bibfnamefont
  {M.}~\bibnamefont {Hatridge}},\ }\bibfield  {title} {\enquote {\bibinfo
  {title} {{Josephson parametric converter saturation and higher order
  effects}},}\ }\href {\doibase 10.1063/1.5003032} {\bibfield  {journal}
  {\bibinfo  {journal} {Applied Physics Letters}\ }\textbf {\bibinfo {volume}
  {111}},\ \bibinfo {pages} {202603} (\bibinfo {year} {2017})}\BibitemShut
  {NoStop}%
\bibitem [{\citenamefont {Frattini}\ \emph {et~al.}(2018)\citenamefont
  {Frattini}, \citenamefont {Sivak}, \citenamefont {Lingenfelter},
  \citenamefont {Shankar},\ and\ \citenamefont {Devoret}}]{SPA}%
  \BibitemOpen
  \bibfield  {author} {\bibinfo {author} {\bibfnamefont {N~E}\ \bibnamefont
  {Frattini}}, \bibinfo {author} {\bibfnamefont {V~V}\ \bibnamefont {Sivak}},
  \bibinfo {author} {\bibfnamefont {A}~\bibnamefont {Lingenfelter}}, \bibinfo
  {author} {\bibfnamefont {S}~\bibnamefont {Shankar}}, \ and\ \bibinfo {author}
  {\bibfnamefont {M~H}\ \bibnamefont {Devoret}},\ }\bibfield  {title} {\enquote
  {\bibinfo {title} {{Optimizing the Nonlinearity and Dissipation of a SNAIL
  Parametric Amplifier for Dynamic Range}},}\ }\href {\doibase
  10.1103/PhysRevApplied.10.054020} {\bibfield  {journal} {\bibinfo  {journal}
  {Physical Review Applied}\ }\textbf {\bibinfo {volume} {10}},\ \bibinfo
  {pages} {54020} (\bibinfo {year} {2018})}\BibitemShut {NoStop}%
\bibitem [{\citenamefont {Naaman}\ \emph {et~al.}(2019)\citenamefont {Naaman},
  \citenamefont {Ferguson}, \citenamefont {Marakov}, \citenamefont {Khalil},
  \citenamefont {Koehl},\ and\ \citenamefont {Epstein}}]{Naaman2019}%
  \BibitemOpen
  \bibfield  {author} {\bibinfo {author} {\bibfnamefont {O.}~\bibnamefont
  {Naaman}}, \bibinfo {author} {\bibfnamefont {D.~G.}\ \bibnamefont
  {Ferguson}}, \bibinfo {author} {\bibfnamefont {A.}~\bibnamefont {Marakov}},
  \bibinfo {author} {\bibfnamefont {M.}~\bibnamefont {Khalil}}, \bibinfo
  {author} {\bibfnamefont {W.~F.}\ \bibnamefont {Koehl}}, \ and\ \bibinfo
  {author} {\bibfnamefont {R.~J.}\ \bibnamefont {Epstein}},\ }\bibfield
  {title} {\enquote {\bibinfo {title} {High saturation power josephson
  parametric amplifier with ghz bandwidth},}\ }in\ \href {\doibase
  10.1109/MWSYM.2019.8701068} {\emph {\bibinfo {booktitle} {2019 IEEE MTT-S
  International Microwave Symposium (IMS)}}}\ (\bibinfo {year} {2019})\ pp.\
  \bibinfo {pages} {259--262}\BibitemShut {NoStop}%
\bibitem [{\citenamefont {Planat}\ \emph {et~al.}(2019)\citenamefont {Planat},
  \citenamefont {Dassonneville}, \citenamefont {Mart\'{\i}nez}, \citenamefont
  {Foroughi}, \citenamefont {Buisson}, \citenamefont {Hasch-Guichard},
  \citenamefont {Naud}, \citenamefont {Vijay}, \citenamefont {Murch},\ and\
  \citenamefont {Roch}}]{Saturation_Planat}%
  \BibitemOpen
  \bibfield  {author} {\bibinfo {author} {\bibfnamefont {Luca}\ \bibnamefont
  {Planat}}, \bibinfo {author} {\bibfnamefont {R\'emy}\ \bibnamefont
  {Dassonneville}}, \bibinfo {author} {\bibfnamefont {Javier~Puertas}\
  \bibnamefont {Mart\'{\i}nez}}, \bibinfo {author} {\bibfnamefont {Farshad}\
  \bibnamefont {Foroughi}}, \bibinfo {author} {\bibfnamefont {Olivier}\
  \bibnamefont {Buisson}}, \bibinfo {author} {\bibfnamefont {Wiebke}\
  \bibnamefont {Hasch-Guichard}}, \bibinfo {author} {\bibfnamefont {C\'ecile}\
  \bibnamefont {Naud}}, \bibinfo {author} {\bibfnamefont {R.}~\bibnamefont
  {Vijay}}, \bibinfo {author} {\bibfnamefont {Kater}\ \bibnamefont {Murch}}, \
  and\ \bibinfo {author} {\bibfnamefont {Nicolas}\ \bibnamefont {Roch}},\
  }\bibfield  {title} {\enquote {\bibinfo {title} {Understanding the saturation
  power of josephson parametric amplifiers made from squid arrays},}\ }\href
  {\doibase 10.1103/PhysRevApplied.11.034014} {\bibfield  {journal} {\bibinfo
  {journal} {Phys. Rev. Appl.}\ }\textbf {\bibinfo {volume} {11}},\ \bibinfo
  {pages} {034014} (\bibinfo {year} {2019})}\BibitemShut {NoStop}%
\bibitem [{\citenamefont {Kaufman}\ \emph {et~al.}(2023)\citenamefont
  {Kaufman}, \citenamefont {White}, \citenamefont {Dykman}, \citenamefont
  {Iorio}, \citenamefont {Sterling}, \citenamefont {Hong}, \citenamefont
  {Opremcak}, \citenamefont {Bengtsson}, \citenamefont {Faoro}, \citenamefont
  {Bardin}, \citenamefont {Burger}, \citenamefont {Gasca},\ and\ \citenamefont
  {Naaman}}]{kaufman_Chebyshev2023}%
  \BibitemOpen
  \bibfield  {author} {\bibinfo {author} {\bibfnamefont {Ryan}\ \bibnamefont
  {Kaufman}}, \bibinfo {author} {\bibfnamefont {Theodore}\ \bibnamefont
  {White}}, \bibinfo {author} {\bibfnamefont {Mark~I.}\ \bibnamefont {Dykman}},
  \bibinfo {author} {\bibfnamefont {Andrea}\ \bibnamefont {Iorio}}, \bibinfo
  {author} {\bibfnamefont {George}\ \bibnamefont {Sterling}}, \bibinfo {author}
  {\bibfnamefont {Sabrina}\ \bibnamefont {Hong}}, \bibinfo {author}
  {\bibfnamefont {Alex}\ \bibnamefont {Opremcak}}, \bibinfo {author}
  {\bibfnamefont {Andreas}\ \bibnamefont {Bengtsson}}, \bibinfo {author}
  {\bibfnamefont {Lara}\ \bibnamefont {Faoro}}, \bibinfo {author}
  {\bibfnamefont {Joseph~C.}\ \bibnamefont {Bardin}}, \bibinfo {author}
  {\bibfnamefont {Tim}\ \bibnamefont {Burger}}, \bibinfo {author}
  {\bibfnamefont {Robert}\ \bibnamefont {Gasca}}, \ and\ \bibinfo {author}
  {\bibfnamefont {Ofer}\ \bibnamefont {Naaman}},\ }\bibfield  {title} {\enquote
  {\bibinfo {title} {Josephson parametric amplifier with chebyshev gain profile
  and high saturation},}\ }\href {\doibase 10.1103/PhysRevApplied.20.054058}
  {\bibfield  {journal} {\bibinfo  {journal} {Phys. Rev. Appl.}\ }\textbf
  {\bibinfo {volume} {20}},\ \bibinfo {pages} {054058} (\bibinfo {year}
  {2023})}\BibitemShut {NoStop}%
\bibitem [{\citenamefont {Kaufman}\ \emph {et~al.}(2024)\citenamefont
  {Kaufman}, \citenamefont {Liu}, \citenamefont {Cicak}, \citenamefont
  {Mesits}, \citenamefont {Xia}, \citenamefont {Zhou}, \citenamefont {Nowicki},
  \citenamefont {Aumentado}, \citenamefont {Pekker},\ and\ \citenamefont
  {Hatridge}}]{kaufman_SimpleHighSaturation2024}%
  \BibitemOpen
  \bibfield  {author} {\bibinfo {author} {\bibfnamefont {Ryan}\ \bibnamefont
  {Kaufman}}, \bibinfo {author} {\bibfnamefont {Chenxu}\ \bibnamefont {Liu}},
  \bibinfo {author} {\bibfnamefont {Katarina}\ \bibnamefont {Cicak}}, \bibinfo
  {author} {\bibfnamefont {Boris}\ \bibnamefont {Mesits}}, \bibinfo {author}
  {\bibfnamefont {Mingkang}\ \bibnamefont {Xia}}, \bibinfo {author}
  {\bibfnamefont {Chao}\ \bibnamefont {Zhou}}, \bibinfo {author} {\bibfnamefont
  {Maria}\ \bibnamefont {Nowicki}}, \bibinfo {author} {\bibfnamefont {José}\
  \bibnamefont {Aumentado}}, \bibinfo {author} {\bibfnamefont {David}\
  \bibnamefont {Pekker}}, \ and\ \bibinfo {author} {\bibfnamefont {Michael}\
  \bibnamefont {Hatridge}},\ }\href@noop {} {\enquote {\bibinfo {title}
  {Simple, high saturation power, quantum-limited, rf squid array-based
  josephson parametric amplifiers},}\ } (\bibinfo {year} {2024}),\ \bibinfo
  {note} {\url{https://arxiv.org/abs/2402.19435}}\BibitemShut {NoStop}%
\bibitem [{\citenamefont {Kundu}\ \emph {et~al.}(2019)\citenamefont {Kundu},
  \citenamefont {Gheeraert}, \citenamefont {Hazra}, \citenamefont {Roy},
  \citenamefont {Salunkhe}, \citenamefont {Patankar},\ and\ \citenamefont
  {Vijay}}]{Kundu2019}%
  \BibitemOpen
  \bibfield  {author} {\bibinfo {author} {\bibfnamefont {Suman}\ \bibnamefont
  {Kundu}}, \bibinfo {author} {\bibfnamefont {Nicolas}\ \bibnamefont
  {Gheeraert}}, \bibinfo {author} {\bibfnamefont {Sumeru}\ \bibnamefont
  {Hazra}}, \bibinfo {author} {\bibfnamefont {Tanay}\ \bibnamefont {Roy}},
  \bibinfo {author} {\bibfnamefont {Kishor~V.}\ \bibnamefont {Salunkhe}},
  \bibinfo {author} {\bibfnamefont {Meghan~P.}\ \bibnamefont {Patankar}}, \
  and\ \bibinfo {author} {\bibfnamefont {R.}~\bibnamefont {Vijay}},\ }\bibfield
   {title} {\enquote {\bibinfo {title} {{Multiplexed readout of four qubits in
  3D circuit QED architecture using a broadband Josephson parametric
  amplifier}},}\ }\href {\doibase 10.1063/1.5089729} {\bibfield  {journal}
  {\bibinfo  {journal} {Applied Physics Letters}\ }\textbf {\bibinfo {volume}
  {114}},\ \bibinfo {pages} {172601} (\bibinfo {year} {2019})}\BibitemShut
  {NoStop}%
\bibitem [{\citenamefont {White}\ \emph {et~al.}(2023)\citenamefont {White},
  \citenamefont {Opremcak}, \citenamefont {Sterling}, \citenamefont {Korotkov},
  \citenamefont {Sank}, \citenamefont {Acharya}, \citenamefont {Ansmann},
  \citenamefont {Arute}, \citenamefont {Arya}, \citenamefont {Bardin},
  \citenamefont {Bengtsson}, \citenamefont {Bourassa}, \citenamefont {Bovaird},
  \citenamefont {Brill}, \citenamefont {Buckley}, \citenamefont {Buell},
  \citenamefont {Burger}, \citenamefont {Burkett}, \citenamefont {Bushnell},
  \citenamefont {Chen}, \citenamefont {Chiaro}, \citenamefont {Cogan},
  \citenamefont {Collins}, \citenamefont {Crook}, \citenamefont {Curtin},
  \citenamefont {Demura}, \citenamefont {Dunsworth}, \citenamefont {Erickson},
  \citenamefont {Fatemi}, \citenamefont {Flores-Burgos}, \citenamefont
  {Forati}, \citenamefont {Foxen}, \citenamefont {Giang}, \citenamefont
  {Giustina}, \citenamefont {Dau}, \citenamefont {Hamilton}, \citenamefont
  {Harrington}, \citenamefont {Hilton}, \citenamefont {Hoffmann}, \citenamefont
  {Hong}, \citenamefont {Huang}, \citenamefont {Huff}, \citenamefont {Iveland},
  \citenamefont {Jeffrey}, \citenamefont {Kieferová}, \citenamefont {Kim},
  \citenamefont {Klimov}, \citenamefont {Kostritsa}, \citenamefont
  {Kreikebaum}, \citenamefont {Landhuis}, \citenamefont {Laptev}, \citenamefont
  {Laws}, \citenamefont {Lee}, \citenamefont {Lester}, \citenamefont {Lill},
  \citenamefont {Liu}, \citenamefont {Locharla}, \citenamefont {Lucero},
  \citenamefont {McCourt}, \citenamefont {McEwen}, \citenamefont {Mi},
  \citenamefont {Miao}, \citenamefont {Montazeri}, \citenamefont {Morvan},
  \citenamefont {Neeley}, \citenamefont {Neill}, \citenamefont {Nersisyan},
  \citenamefont {Ng}, \citenamefont {Nguyen}, \citenamefont {Nguyen},
  \citenamefont {Potter}, \citenamefont {Quintana}, \citenamefont {Roushan},
  \citenamefont {Sankaragomathi}, \citenamefont {Satzinger}, \citenamefont
  {Schuster}, \citenamefont {Shearn}, \citenamefont {Shorter}, \citenamefont
  {Shvarts}, \citenamefont {Skruzny}, \citenamefont {Smith}, \citenamefont
  {Szalay}, \citenamefont {Torres}, \citenamefont {Woo}, \citenamefont {Yao},
  \citenamefont {Yeh}, \citenamefont {Yoo}, \citenamefont {Young},
  \citenamefont {Zhu}, \citenamefont {Zobrist}, \citenamefont {Chen},
  \citenamefont {Megrant}, \citenamefont {Kelly},\ and\ \citenamefont
  {Naaman}}]{whiteReadoutQuantumProcessor2023}%
  \BibitemOpen
  \bibfield  {author} {\bibinfo {author} {\bibfnamefont {T.~C.}\ \bibnamefont
  {White}}, \bibinfo {author} {\bibfnamefont {Alex}\ \bibnamefont {Opremcak}},
  \bibinfo {author} {\bibfnamefont {George}\ \bibnamefont {Sterling}}, \bibinfo
  {author} {\bibfnamefont {Alexander}\ \bibnamefont {Korotkov}}, \bibinfo
  {author} {\bibfnamefont {Daniel}\ \bibnamefont {Sank}}, \bibinfo {author}
  {\bibfnamefont {Rajeev}\ \bibnamefont {Acharya}}, \bibinfo {author}
  {\bibfnamefont {Markus}\ \bibnamefont {Ansmann}}, \bibinfo {author}
  {\bibfnamefont {Frank}\ \bibnamefont {Arute}}, \bibinfo {author}
  {\bibfnamefont {Kunal}\ \bibnamefont {Arya}}, \bibinfo {author}
  {\bibfnamefont {Joseph~C.}\ \bibnamefont {Bardin}}, \bibinfo {author}
  {\bibfnamefont {Andreas}\ \bibnamefont {Bengtsson}}, \bibinfo {author}
  {\bibfnamefont {Alexandre}\ \bibnamefont {Bourassa}}, \bibinfo {author}
  {\bibfnamefont {Jenna}\ \bibnamefont {Bovaird}}, \bibinfo {author}
  {\bibfnamefont {Leon}\ \bibnamefont {Brill}}, \bibinfo {author}
  {\bibfnamefont {Bob~B.}\ \bibnamefont {Buckley}}, \bibinfo {author}
  {\bibfnamefont {David~A.}\ \bibnamefont {Buell}}, \bibinfo {author}
  {\bibfnamefont {Tim}\ \bibnamefont {Burger}}, \bibinfo {author}
  {\bibfnamefont {Brian}\ \bibnamefont {Burkett}}, \bibinfo {author}
  {\bibfnamefont {Nicholas}\ \bibnamefont {Bushnell}}, \bibinfo {author}
  {\bibfnamefont {Zijun}\ \bibnamefont {Chen}}, \bibinfo {author}
  {\bibfnamefont {Ben}\ \bibnamefont {Chiaro}}, \bibinfo {author}
  {\bibfnamefont {Josh}\ \bibnamefont {Cogan}}, \bibinfo {author}
  {\bibfnamefont {Roberto}\ \bibnamefont {Collins}}, \bibinfo {author}
  {\bibfnamefont {Alexander~L.}\ \bibnamefont {Crook}}, \bibinfo {author}
  {\bibfnamefont {Ben}\ \bibnamefont {Curtin}}, \bibinfo {author}
  {\bibfnamefont {Sean}\ \bibnamefont {Demura}}, \bibinfo {author}
  {\bibfnamefont {Andrew}\ \bibnamefont {Dunsworth}}, \bibinfo {author}
  {\bibfnamefont {Catherine}\ \bibnamefont {Erickson}}, \bibinfo {author}
  {\bibfnamefont {Reza}\ \bibnamefont {Fatemi}}, \bibinfo {author}
  {\bibfnamefont {Leslie}\ \bibnamefont {Flores-Burgos}}, \bibinfo {author}
  {\bibfnamefont {Ebrahim}\ \bibnamefont {Forati}}, \bibinfo {author}
  {\bibfnamefont {Brooks}\ \bibnamefont {Foxen}}, \bibinfo {author}
  {\bibfnamefont {William}\ \bibnamefont {Giang}}, \bibinfo {author}
  {\bibfnamefont {Marissa}\ \bibnamefont {Giustina}}, \bibinfo {author}
  {\bibfnamefont {Alejandro~Grajales}\ \bibnamefont {Dau}}, \bibinfo {author}
  {\bibfnamefont {Michael~C.}\ \bibnamefont {Hamilton}}, \bibinfo {author}
  {\bibfnamefont {Sean~D.}\ \bibnamefont {Harrington}}, \bibinfo {author}
  {\bibfnamefont {Jeremy}\ \bibnamefont {Hilton}}, \bibinfo {author}
  {\bibfnamefont {Markus}\ \bibnamefont {Hoffmann}}, \bibinfo {author}
  {\bibfnamefont {Sabrina}\ \bibnamefont {Hong}}, \bibinfo {author}
  {\bibfnamefont {Trent}\ \bibnamefont {Huang}}, \bibinfo {author}
  {\bibfnamefont {Ashley}\ \bibnamefont {Huff}}, \bibinfo {author}
  {\bibfnamefont {Justin}\ \bibnamefont {Iveland}}, \bibinfo {author}
  {\bibfnamefont {Evan}\ \bibnamefont {Jeffrey}}, \bibinfo {author}
  {\bibfnamefont {Márika}\ \bibnamefont {Kieferová}}, \bibinfo {author}
  {\bibfnamefont {Seon}\ \bibnamefont {Kim}}, \bibinfo {author} {\bibfnamefont
  {Paul~V.}\ \bibnamefont {Klimov}}, \bibinfo {author} {\bibfnamefont {Fedor}\
  \bibnamefont {Kostritsa}}, \bibinfo {author} {\bibfnamefont {John~Mark}\
  \bibnamefont {Kreikebaum}}, \bibinfo {author} {\bibfnamefont {David}\
  \bibnamefont {Landhuis}}, \bibinfo {author} {\bibfnamefont {Pavel}\
  \bibnamefont {Laptev}}, \bibinfo {author} {\bibfnamefont {Lily}\ \bibnamefont
  {Laws}}, \bibinfo {author} {\bibfnamefont {Kenny}\ \bibnamefont {Lee}},
  \bibinfo {author} {\bibfnamefont {Brian~J.}\ \bibnamefont {Lester}}, \bibinfo
  {author} {\bibfnamefont {Alexander}\ \bibnamefont {Lill}}, \bibinfo {author}
  {\bibfnamefont {Wayne}\ \bibnamefont {Liu}}, \bibinfo {author} {\bibfnamefont
  {Aditya}\ \bibnamefont {Locharla}}, \bibinfo {author} {\bibfnamefont {Erik}\
  \bibnamefont {Lucero}}, \bibinfo {author} {\bibfnamefont {Trevor}\
  \bibnamefont {McCourt}}, \bibinfo {author} {\bibfnamefont {Matt}\
  \bibnamefont {McEwen}}, \bibinfo {author} {\bibfnamefont {Xiao}\ \bibnamefont
  {Mi}}, \bibinfo {author} {\bibfnamefont {Kevin~C.}\ \bibnamefont {Miao}},
  \bibinfo {author} {\bibfnamefont {Shirin}\ \bibnamefont {Montazeri}},
  \bibinfo {author} {\bibfnamefont {Alexis}\ \bibnamefont {Morvan}}, \bibinfo
  {author} {\bibfnamefont {Matthew}\ \bibnamefont {Neeley}}, \bibinfo {author}
  {\bibfnamefont {Charles}\ \bibnamefont {Neill}}, \bibinfo {author}
  {\bibfnamefont {Ani}\ \bibnamefont {Nersisyan}}, \bibinfo {author}
  {\bibfnamefont {Jiun~How}\ \bibnamefont {Ng}}, \bibinfo {author}
  {\bibfnamefont {Anthony}\ \bibnamefont {Nguyen}}, \bibinfo {author}
  {\bibfnamefont {Murray}\ \bibnamefont {Nguyen}}, \bibinfo {author}
  {\bibfnamefont {Rebecca}\ \bibnamefont {Potter}}, \bibinfo {author}
  {\bibfnamefont {Chris}\ \bibnamefont {Quintana}}, \bibinfo {author}
  {\bibfnamefont {Pedram}\ \bibnamefont {Roushan}}, \bibinfo {author}
  {\bibfnamefont {Kannan}\ \bibnamefont {Sankaragomathi}}, \bibinfo {author}
  {\bibfnamefont {Kevin~J.}\ \bibnamefont {Satzinger}}, \bibinfo {author}
  {\bibfnamefont {Christopher}\ \bibnamefont {Schuster}}, \bibinfo {author}
  {\bibfnamefont {Michael~J.}\ \bibnamefont {Shearn}}, \bibinfo {author}
  {\bibfnamefont {Aaron}\ \bibnamefont {Shorter}}, \bibinfo {author}
  {\bibfnamefont {Vladimir}\ \bibnamefont {Shvarts}}, \bibinfo {author}
  {\bibfnamefont {Jindra}\ \bibnamefont {Skruzny}}, \bibinfo {author}
  {\bibfnamefont {W.~Clarke}\ \bibnamefont {Smith}}, \bibinfo {author}
  {\bibfnamefont {Marco}\ \bibnamefont {Szalay}}, \bibinfo {author}
  {\bibfnamefont {Alfredo}\ \bibnamefont {Torres}}, \bibinfo {author}
  {\bibfnamefont {Bryan}\ \bibnamefont {Woo}}, \bibinfo {author} {\bibfnamefont
  {Z.~Jamie}\ \bibnamefont {Yao}}, \bibinfo {author} {\bibfnamefont {Ping}\
  \bibnamefont {Yeh}}, \bibinfo {author} {\bibfnamefont {Juhwan}\ \bibnamefont
  {Yoo}}, \bibinfo {author} {\bibfnamefont {Grayson}\ \bibnamefont {Young}},
  \bibinfo {author} {\bibfnamefont {Ningfeng}\ \bibnamefont {Zhu}}, \bibinfo
  {author} {\bibfnamefont {Nicholas}\ \bibnamefont {Zobrist}}, \bibinfo
  {author} {\bibfnamefont {Yu}~\bibnamefont {Chen}}, \bibinfo {author}
  {\bibfnamefont {Anthony}\ \bibnamefont {Megrant}}, \bibinfo {author}
  {\bibfnamefont {Julian}\ \bibnamefont {Kelly}}, \ and\ \bibinfo {author}
  {\bibfnamefont {Ofer}\ \bibnamefont {Naaman}},\ }\bibfield  {title} {\enquote
  {\bibinfo {title} {Readout of a quantum processor with high dynamic range
  {Josephson} parametric amplifiers},}\ }\href {\doibase 10.1063/5.0127375}
  {\bibfield  {journal} {\bibinfo  {journal} {Applied Physics Letters}\
  }\textbf {\bibinfo {volume} {122}},\ \bibinfo {pages} {014001} (\bibinfo
  {year} {2023})}\BibitemShut {NoStop}%
\bibitem [{\citenamefont {Sivak}\ \emph {et~al.}(2019)\citenamefont {Sivak},
  \citenamefont {Frattini}, \citenamefont {Joshi}, \citenamefont
  {Lingenfelter}, \citenamefont {Shankar},\ and\ \citenamefont
  {Devoret}}]{Kerr_free}%
  \BibitemOpen
  \bibfield  {author} {\bibinfo {author} {\bibfnamefont {V.V.}\ \bibnamefont
  {Sivak}}, \bibinfo {author} {\bibfnamefont {N.E.}\ \bibnamefont {Frattini}},
  \bibinfo {author} {\bibfnamefont {V.R.}\ \bibnamefont {Joshi}}, \bibinfo
  {author} {\bibfnamefont {A.}~\bibnamefont {Lingenfelter}}, \bibinfo {author}
  {\bibfnamefont {S.}~\bibnamefont {Shankar}}, \ and\ \bibinfo {author}
  {\bibfnamefont {M.H.}\ \bibnamefont {Devoret}},\ }\bibfield  {title}
  {\enquote {\bibinfo {title} {Kerr-free three-wave mixing in superconducting
  quantum circuits},}\ }\href {\doibase 10.1103/PhysRevApplied.11.054060}
  {\bibfield  {journal} {\bibinfo  {journal} {Phys. Rev. Appl.}\ }\textbf
  {\bibinfo {volume} {11}},\ \bibinfo {pages} {054060} (\bibinfo {year}
  {2019})}\BibitemShut {NoStop}%
\bibitem [{\citenamefont {Liu}\ \emph {et~al.}(2020)\citenamefont {Liu},
  \citenamefont {Chien}, \citenamefont {Hatridge},\ and\ \citenamefont
  {Pekker}}]{JPC_Hamiltonian_control}%
  \BibitemOpen
  \bibfield  {author} {\bibinfo {author} {\bibfnamefont {Chenxu}\ \bibnamefont
  {Liu}}, \bibinfo {author} {\bibfnamefont {Tzu-Chiao}\ \bibnamefont {Chien}},
  \bibinfo {author} {\bibfnamefont {Michael}\ \bibnamefont {Hatridge}}, \ and\
  \bibinfo {author} {\bibfnamefont {David}\ \bibnamefont {Pekker}},\ }\bibfield
   {title} {\enquote {\bibinfo {title} {Optimizing
  josephson-ring-modulator-based josephson parametric amplifiers via full
  hamiltonian control},}\ }\href {\doibase 10.1103/PhysRevA.101.042323}
  {\bibfield  {journal} {\bibinfo  {journal} {Phys. Rev. A}\ }\textbf {\bibinfo
  {volume} {101}},\ \bibinfo {pages} {042323} (\bibinfo {year}
  {2020})}\BibitemShut {NoStop}%
\bibitem [{\citenamefont {Sivak}\ \emph {et~al.}(2020)\citenamefont {Sivak},
  \citenamefont {Shankar}, \citenamefont {Liu}, \citenamefont {Aumentado},\
  and\ \citenamefont {Devoret}}]{JAMPA}%
  \BibitemOpen
  \bibfield  {author} {\bibinfo {author} {\bibfnamefont {V.~V.}\ \bibnamefont
  {Sivak}}, \bibinfo {author} {\bibfnamefont {S.}~\bibnamefont {Shankar}},
  \bibinfo {author} {\bibfnamefont {G.}~\bibnamefont {Liu}}, \bibinfo {author}
  {\bibfnamefont {J.}~\bibnamefont {Aumentado}}, \ and\ \bibinfo {author}
  {\bibfnamefont {M.~H.}\ \bibnamefont {Devoret}},\ }\bibfield  {title}
  {\enquote {\bibinfo {title} {Josephson array-mode parametric amplifier},}\
  }\href {\doibase 10.1103/PhysRevApplied.13.024014} {\bibfield  {journal}
  {\bibinfo  {journal} {Phys. Rev. Appl.}\ }\textbf {\bibinfo {volume} {13}},\
  \bibinfo {pages} {024014} (\bibinfo {year} {2020})}\BibitemShut {NoStop}%
\bibitem [{\citenamefont {Kamal}\ \emph {et~al.}(2009)\citenamefont {Kamal},
  \citenamefont {Marblestone},\ and\ \citenamefont {Devoret}}]{Double-pump}%
  \BibitemOpen
  \bibfield  {author} {\bibinfo {author} {\bibfnamefont {Archana}\ \bibnamefont
  {Kamal}}, \bibinfo {author} {\bibfnamefont {Adam}\ \bibnamefont
  {Marblestone}}, \ and\ \bibinfo {author} {\bibfnamefont {Michel}\
  \bibnamefont {Devoret}},\ }\bibfield  {title} {\enquote {\bibinfo {title}
  {Signal-to-pump back action and self-oscillation in double-pump josephson
  parametric amplifier},}\ }\href {\doibase 10.1103/PhysRevB.79.184301}
  {\bibfield  {journal} {\bibinfo  {journal} {Phys. Rev. B}\ }\textbf {\bibinfo
  {volume} {79}},\ \bibinfo {pages} {184301} (\bibinfo {year}
  {2009})}\BibitemShut {NoStop}%
\bibitem [{\citenamefont {Yamamoto}\ \emph {et~al.}(2008)\citenamefont
  {Yamamoto}, \citenamefont {Inomata}, \citenamefont {Watanabe}, \citenamefont
  {Matsuba}, \citenamefont {Miyazaki}, \citenamefont {Oliver}, \citenamefont
  {Nakamura},\ and\ \citenamefont {Tsai}}]{Yamamoto2008}%
  \BibitemOpen
  \bibfield  {author} {\bibinfo {author} {\bibfnamefont {T}~\bibnamefont
  {Yamamoto}}, \bibinfo {author} {\bibfnamefont {K}~\bibnamefont {Inomata}},
  \bibinfo {author} {\bibfnamefont {M}~\bibnamefont {Watanabe}}, \bibinfo
  {author} {\bibfnamefont {K}~\bibnamefont {Matsuba}}, \bibinfo {author}
  {\bibfnamefont {T}~\bibnamefont {Miyazaki}}, \bibinfo {author} {\bibfnamefont
  {W~D}\ \bibnamefont {Oliver}}, \bibinfo {author} {\bibfnamefont
  {Y}~\bibnamefont {Nakamura}}, \ and\ \bibinfo {author} {\bibfnamefont {J~S}\
  \bibnamefont {Tsai}},\ }\bibfield  {title} {\enquote {\bibinfo {title}
  {{Flux-driven Josephson parametric amplifier}},}\ }\href {\doibase
  10.1063/1.2964182} {\bibfield  {journal} {\bibinfo  {journal} {Applied
  Physics Letters}\ }\textbf {\bibinfo {volume} {93}},\ \bibinfo {pages}
  {42510} (\bibinfo {year} {2008})}\BibitemShut {NoStop}%
\bibitem [{\citenamefont {Simoen}\ \emph {et~al.}(2015)\citenamefont {Simoen},
  \citenamefont {Chang}, \citenamefont {Krantz}, \citenamefont {Bylander},
  \citenamefont {Wustmann}, \citenamefont {Shumeiko}, \citenamefont {Delsing},\
  and\ \citenamefont {Wilson}}]{Simoen2015}%
  \BibitemOpen
  \bibfield  {author} {\bibinfo {author} {\bibfnamefont {M.}~\bibnamefont
  {Simoen}}, \bibinfo {author} {\bibfnamefont {C.~W.~S.}\ \bibnamefont
  {Chang}}, \bibinfo {author} {\bibfnamefont {P.}~\bibnamefont {Krantz}},
  \bibinfo {author} {\bibfnamefont {Jonas}\ \bibnamefont {Bylander}}, \bibinfo
  {author} {\bibfnamefont {W.}~\bibnamefont {Wustmann}}, \bibinfo {author}
  {\bibfnamefont {V.}~\bibnamefont {Shumeiko}}, \bibinfo {author}
  {\bibfnamefont {P.}~\bibnamefont {Delsing}}, \ and\ \bibinfo {author}
  {\bibfnamefont {C.~M.}\ \bibnamefont {Wilson}},\ }\bibfield  {title}
  {\enquote {\bibinfo {title} {{Characterization of a multimode coplanar
  waveguide parametric amplifier}},}\ }\href {\doibase 10.1063/1.4933265}
  {\bibfield  {journal} {\bibinfo  {journal} {Journal of Applied Physics}\
  }\textbf {\bibinfo {volume} {118}},\ \bibinfo {pages} {154501} (\bibinfo
  {year} {2015})}\BibitemShut {NoStop}%
\bibitem [{\citenamefont {Zhou}\ \emph {et~al.}(2014)\citenamefont {Zhou},
  \citenamefont {Schmitt}, \citenamefont {Bertet}, \citenamefont {Vion},
  \citenamefont {Wustmann}, \citenamefont {Shumeiko},\ and\ \citenamefont
  {Esteve}}]{Zhou2014}%
  \BibitemOpen
  \bibfield  {author} {\bibinfo {author} {\bibfnamefont {X}~\bibnamefont
  {Zhou}}, \bibinfo {author} {\bibfnamefont {V}~\bibnamefont {Schmitt}},
  \bibinfo {author} {\bibfnamefont {P}~\bibnamefont {Bertet}}, \bibinfo
  {author} {\bibfnamefont {D}~\bibnamefont {Vion}}, \bibinfo {author}
  {\bibfnamefont {W}~\bibnamefont {Wustmann}}, \bibinfo {author} {\bibfnamefont
  {V}~\bibnamefont {Shumeiko}}, \ and\ \bibinfo {author} {\bibfnamefont
  {D}~\bibnamefont {Esteve}},\ }\bibfield  {title} {\enquote {\bibinfo {title}
  {{High-gain weakly nonlinear flux-modulated Josephson parametric amplifier
  using a SQUID array}},}\ }\href {\doibase 10.1103/PhysRevB.89.214517}
  {\bibfield  {journal} {\bibinfo  {journal} {Physical Review B}\ }\textbf
  {\bibinfo {volume} {89}},\ \bibinfo {pages} {214517} (\bibinfo {year}
  {2014})}\BibitemShut {NoStop}%
\bibitem [{\citenamefont {Elo}\ \emph {et~al.}(2019)\citenamefont {Elo},
  \citenamefont {Abhilash}, \citenamefont {Perelshtein}, \citenamefont {Lilja},
  \citenamefont {Korostylev},\ and\ \citenamefont {Hakonen}}]{Elo2019}%
  \BibitemOpen
  \bibfield  {author} {\bibinfo {author} {\bibfnamefont {T.}~\bibnamefont
  {Elo}}, \bibinfo {author} {\bibfnamefont {T.~S.}\ \bibnamefont {Abhilash}},
  \bibinfo {author} {\bibfnamefont {M.~R.}\ \bibnamefont {Perelshtein}},
  \bibinfo {author} {\bibfnamefont {I.}~\bibnamefont {Lilja}}, \bibinfo
  {author} {\bibfnamefont {E.~V.}\ \bibnamefont {Korostylev}}, \ and\ \bibinfo
  {author} {\bibfnamefont {P.~J.}\ \bibnamefont {Hakonen}},\ }\bibfield
  {title} {\enquote {\bibinfo {title} {{Broadband lumped-element Josephson
  parametric amplifier with single-step lithography}},}\ }\href {\doibase
  10.1063/1.5086091} {\bibfield  {journal} {\bibinfo  {journal} {Applied
  Physics Letters}\ }\textbf {\bibinfo {volume} {114}},\ \bibinfo {pages}
  {152601} (\bibinfo {year} {2019})}\BibitemShut {NoStop}%
\bibitem [{\citenamefont {Mutus}\ \emph {et~al.}(2013)\citenamefont {Mutus},
  \citenamefont {White}, \citenamefont {Jeffrey}, \citenamefont {Sank},
  \citenamefont {Barends}, \citenamefont {Bochmann}, \citenamefont {Chen},
  \citenamefont {Chen}, \citenamefont {Chiaro}, \citenamefont {Dunsworth},
  \citenamefont {Kelly}, \citenamefont {Megrant}, \citenamefont {Neill},
  \citenamefont {O'Malley}, \citenamefont {Roushan}, \citenamefont
  {Vainsencher}, \citenamefont {Wenner}, \citenamefont {Siddiqi}, \citenamefont
  {Vijay}, \citenamefont {Cleland},\ and\ \citenamefont
  {Martinis}}]{Mutus2013}%
  \BibitemOpen
  \bibfield  {author} {\bibinfo {author} {\bibfnamefont {J.~Y.}\ \bibnamefont
  {Mutus}}, \bibinfo {author} {\bibfnamefont {T.~C.}\ \bibnamefont {White}},
  \bibinfo {author} {\bibfnamefont {E.}~\bibnamefont {Jeffrey}}, \bibinfo
  {author} {\bibfnamefont {D.}~\bibnamefont {Sank}}, \bibinfo {author}
  {\bibfnamefont {R.}~\bibnamefont {Barends}}, \bibinfo {author} {\bibfnamefont
  {J.}~\bibnamefont {Bochmann}}, \bibinfo {author} {\bibfnamefont
  {Yu}~\bibnamefont {Chen}}, \bibinfo {author} {\bibfnamefont {Z.}~\bibnamefont
  {Chen}}, \bibinfo {author} {\bibfnamefont {B.}~\bibnamefont {Chiaro}},
  \bibinfo {author} {\bibfnamefont {A.}~\bibnamefont {Dunsworth}}, \bibinfo
  {author} {\bibfnamefont {J.}~\bibnamefont {Kelly}}, \bibinfo {author}
  {\bibfnamefont {A.}~\bibnamefont {Megrant}}, \bibinfo {author} {\bibfnamefont
  {C.}~\bibnamefont {Neill}}, \bibinfo {author} {\bibfnamefont {P.~J.J.}\
  \bibnamefont {O'Malley}}, \bibinfo {author} {\bibfnamefont {P.}~\bibnamefont
  {Roushan}}, \bibinfo {author} {\bibfnamefont {A.}~\bibnamefont
  {Vainsencher}}, \bibinfo {author} {\bibfnamefont {J.}~\bibnamefont {Wenner}},
  \bibinfo {author} {\bibfnamefont {I.}~\bibnamefont {Siddiqi}}, \bibinfo
  {author} {\bibfnamefont {R.}~\bibnamefont {Vijay}}, \bibinfo {author}
  {\bibfnamefont {A.~N.}\ \bibnamefont {Cleland}}, \ and\ \bibinfo {author}
  {\bibfnamefont {John~M.}\ \bibnamefont {Martinis}},\ }\bibfield  {title}
  {\enquote {\bibinfo {title} {{Design and characterization of a lumped element
  single-ended superconducting microwave parametric amplifier with on-chip flux
  bias line}},}\ }\href {https://doi.org/10.1063/1.4821136} {\bibfield
  {journal} {\bibinfo  {journal} {Applied Physics Letters}\ } (\bibinfo {year}
  {2013})}\BibitemShut {NoStop}%
\bibitem [{\citenamefont {{Urade}}\ \emph {et~al.}(2021)\citenamefont
  {{Urade}}, \citenamefont {{Zuo}}, \citenamefont {{Baba}}, \citenamefont
  {{Chang}}, \citenamefont {{Nittoh}}, \citenamefont {{Inomata}}, \citenamefont
  {{Lin}}, \citenamefont {{Yamamoto}},\ and\ \citenamefont
  {{Nakamura}}}]{Urade_MM2021}%
  \BibitemOpen
  \bibfield  {author} {\bibinfo {author} {\bibfnamefont {Yoshiro}\ \bibnamefont
  {{Urade}}}, \bibinfo {author} {\bibfnamefont {Kun}\ \bibnamefont {{Zuo}}},
  \bibinfo {author} {\bibfnamefont {Syotaro}\ \bibnamefont {{Baba}}}, \bibinfo
  {author} {\bibfnamefont {C.~W.~Sandbo}\ \bibnamefont {{Chang}}}, \bibinfo
  {author} {\bibfnamefont {Koh-Ichi}\ \bibnamefont {{Nittoh}}}, \bibinfo
  {author} {\bibfnamefont {Kunihiro}\ \bibnamefont {{Inomata}}}, \bibinfo
  {author} {\bibfnamefont {Zhirong}\ \bibnamefont {{Lin}}}, \bibinfo {author}
  {\bibfnamefont {Tsuyoshi}\ \bibnamefont {{Yamamoto}}}, \ and\ \bibinfo
  {author} {\bibfnamefont {Yasunobu}\ \bibnamefont {{Nakamura}}},\ }\href@noop
  {} {\enquote {\bibinfo {title} {{Flux-driven impedance-matched Josephson
  parametric amplifier with improved pump efficiency}},}\ }\bibinfo
  {howpublished} {Bulletin of the American Physical Society} (\bibinfo {year}
  {2021}),\ \bibinfo {note} {a28.10}\BibitemShut {NoStop}%
\bibitem [{\citenamefont {Hougland}\ \emph {et~al.}(2025)\citenamefont
  {Hougland}, \citenamefont {Li}, \citenamefont {Kaufman}, \citenamefont
  {Mesits}, \citenamefont {Mong}, \citenamefont {Hatridge},\ and\ \citenamefont
  {Pekker}}]{hougland2024pumpefficient}%
  \BibitemOpen
  \bibfield  {author} {\bibinfo {author} {\bibfnamefont {Nicholas~M.}\
  \bibnamefont {Hougland}}, \bibinfo {author} {\bibfnamefont {Zhuan}\
  \bibnamefont {Li}}, \bibinfo {author} {\bibfnamefont {Ryan}\ \bibnamefont
  {Kaufman}}, \bibinfo {author} {\bibfnamefont {Boris}\ \bibnamefont {Mesits}},
  \bibinfo {author} {\bibfnamefont {Roger S.~K.}\ \bibnamefont {Mong}},
  \bibinfo {author} {\bibfnamefont {Michael}\ \bibnamefont {Hatridge}}, \ and\
  \bibinfo {author} {\bibfnamefont {David}\ \bibnamefont {Pekker}},\ }\bibfield
   {title} {\enquote {\bibinfo {title} {Pump-efficient josephson parametric
  amplifiers with high saturation power},}\ }\href {\doibase
  10.1103/PhysRevA.111.022611} {\bibfield  {journal} {\bibinfo  {journal}
  {Phys. Rev. A}\ }\textbf {\bibinfo {volume} {111}},\ \bibinfo {pages}
  {022611} (\bibinfo {year} {2025})}\BibitemShut {NoStop}%
\bibitem [{\citenamefont {Schackert}\ \emph {et~al.}(2013)\citenamefont
  {Schackert}, \citenamefont {Roy}, \citenamefont {Hatridge}, \citenamefont
  {Devoret},\ and\ \citenamefont {Stone}}]{Schackert2013}%
  \BibitemOpen
  \bibfield  {author} {\bibinfo {author} {\bibfnamefont {Flavius}\ \bibnamefont
  {Schackert}}, \bibinfo {author} {\bibfnamefont {Ananda}\ \bibnamefont {Roy}},
  \bibinfo {author} {\bibfnamefont {Michael}\ \bibnamefont {Hatridge}},
  \bibinfo {author} {\bibfnamefont {Michel~H.}\ \bibnamefont {Devoret}}, \ and\
  \bibinfo {author} {\bibfnamefont {A.~Douglas}\ \bibnamefont {Stone}},\
  }\bibfield  {title} {\enquote {\bibinfo {title} {Three-wave mixing with three
  incoming waves: Signal-idler coherent attenuation and gain enhancement in a
  parametric amplifier},}\ }\href {\doibase 10.1103/PhysRevLett.111.073903}
  {\bibfield  {journal} {\bibinfo  {journal} {Phys. Rev. Lett.}\ }\textbf
  {\bibinfo {volume} {111}},\ \bibinfo {pages} {073903} (\bibinfo {year}
  {2013})}\BibitemShut {NoStop}%
\bibitem [{\citenamefont {Sundqvist}\ \emph {et~al.}(2013)\citenamefont
  {Sundqvist}, \citenamefont {Kintaş}, \citenamefont {Simoen}, \citenamefont
  {Krantz}, \citenamefont {Sandberg}, \citenamefont {Wilson},\ and\
  \citenamefont {Delsing}}]{Pumpistor}%
  \BibitemOpen
  \bibfield  {author} {\bibinfo {author} {\bibfnamefont {K.~M.}\ \bibnamefont
  {Sundqvist}}, \bibinfo {author} {\bibfnamefont {S.}~\bibnamefont {Kintaş}},
  \bibinfo {author} {\bibfnamefont {M.}~\bibnamefont {Simoen}}, \bibinfo
  {author} {\bibfnamefont {P.}~\bibnamefont {Krantz}}, \bibinfo {author}
  {\bibfnamefont {M.}~\bibnamefont {Sandberg}}, \bibinfo {author}
  {\bibfnamefont {C.~M.}\ \bibnamefont {Wilson}}, \ and\ \bibinfo {author}
  {\bibfnamefont {P.}~\bibnamefont {Delsing}},\ }\bibfield  {title} {\enquote
  {\bibinfo {title} {The pumpistor: A linearized model of a flux-pumped
  superconducting quantum interference device for use as a negative-resistance
  parametric amplifier},}\ }\href {\doibase 10.1063/1.4819881} {\bibfield
  {journal} {\bibinfo  {journal} {Applied Physics Letters}\ }\textbf {\bibinfo
  {volume} {103}},\ \bibinfo {pages} {102603} (\bibinfo {year}
  {2013})}\BibitemShut {NoStop}%
\bibitem [{\citenamefont {Sundqvist}\ and\ \citenamefont
  {Delsing}(2014)}]{Sundqvist2014}%
  \BibitemOpen
  \bibfield  {author} {\bibinfo {author} {\bibfnamefont {Kyle~M}\ \bibnamefont
  {Sundqvist}}\ and\ \bibinfo {author} {\bibfnamefont {Per}\ \bibnamefont
  {Delsing}},\ }\bibfield  {title} {\enquote {\bibinfo {title}
  {{Negative-resistance models for parametrically flux-pumped superconducting
  quantum interference devices}},}\ }\href {\doibase 10.1140/epjqt6} {\bibfield
   {journal} {\bibinfo  {journal} {EPJ Quantum Technology}\ }\textbf {\bibinfo
  {volume} {1}},\ \bibinfo {pages} {6} (\bibinfo {year} {2014})}\BibitemShut
  {NoStop}%
\bibitem [{\citenamefont {Naaman}\ and\ \citenamefont
  {Aumentado}(2022)}]{naaman2021synthesis}%
  \BibitemOpen
  \bibfield  {author} {\bibinfo {author} {\bibfnamefont {Ofer}\ \bibnamefont
  {Naaman}}\ and\ \bibinfo {author} {\bibfnamefont {Jos\'e}\ \bibnamefont
  {Aumentado}},\ }\bibfield  {title} {\enquote {\bibinfo {title} {Synthesis of
  parametrically coupled networks},}\ }\href {\doibase
  10.1103/PRXQuantum.3.020201} {\bibfield  {journal} {\bibinfo  {journal} {PRX
  Quantum}\ }\textbf {\bibinfo {volume} {3}},\ \bibinfo {pages} {020201}
  (\bibinfo {year} {2022})}\BibitemShut {NoStop}%
\bibitem [{\citenamefont {Frattini}\ \emph {et~al.}(2017)\citenamefont
  {Frattini}, \citenamefont {Vool}, \citenamefont {Shankar}, \citenamefont
  {Narla}, \citenamefont {Sliwa},\ and\ \citenamefont {Devoret}}]{SNAIL}%
  \BibitemOpen
  \bibfield  {author} {\bibinfo {author} {\bibfnamefont {N.~E.}\ \bibnamefont
  {Frattini}}, \bibinfo {author} {\bibfnamefont {U.}~\bibnamefont {Vool}},
  \bibinfo {author} {\bibfnamefont {S.}~\bibnamefont {Shankar}}, \bibinfo
  {author} {\bibfnamefont {A.}~\bibnamefont {Narla}}, \bibinfo {author}
  {\bibfnamefont {K.~M.}\ \bibnamefont {Sliwa}}, \ and\ \bibinfo {author}
  {\bibfnamefont {M.~H.}\ \bibnamefont {Devoret}},\ }\bibfield  {title}
  {\enquote {\bibinfo {title} {{3-wave mixing Josephson dipole element}},}\
  }\href {\doibase 10.1063/1.4984142} {\bibfield  {journal} {\bibinfo
  {journal} {Applied Physics Letters}\ }\textbf {\bibinfo {volume} {110}},\
  \bibinfo {pages} {222603} (\bibinfo {year} {2017})}\BibitemShut {NoStop}%
\bibitem [{\citenamefont {Pozar}(2005)}]{Pozar}%
  \BibitemOpen
  \bibfield  {author} {\bibinfo {author} {\bibfnamefont {David~M}\ \bibnamefont
  {Pozar}},\ }\href {https://cds.cern.ch/record/882338} {\emph {\bibinfo
  {title} {{Microwave engineering; 3rd ed.}}}}\ (\bibinfo  {publisher}
  {Wiley},\ \bibinfo {address} {Hoboken, NJ},\ \bibinfo {year}
  {2005})\BibitemShut {NoStop}%
\bibitem [{\citenamefont {Matthaei}\ \emph {et~al.}(1980)\citenamefont
  {Matthaei}, \citenamefont {Jones},\ and\ \citenamefont
  {Young}}]{matthaeiMicrowaveFiltersImpedanceMatching1980}%
  \BibitemOpen
  \bibfield  {author} {\bibinfo {author} {\bibfnamefont {G.}~\bibnamefont
  {Matthaei}}, \bibinfo {author} {\bibfnamefont {E.~M.~T.}\ \bibnamefont
  {Jones}}, \ and\ \bibinfo {author} {\bibfnamefont {L.}~\bibnamefont
  {Young}},\ }\href@noop {} {\emph {\bibinfo {title} {Microwave {Filters},
  {Impedance}-{Matching} {Networks}, and {Coupling} {Structures}}}}\ (\bibinfo
  {publisher} {Artech House},\ \bibinfo {address} {Norwood, Mass},\ \bibinfo
  {year} {1980})\BibitemShut {NoStop}%
\bibitem [{\citenamefont {Malnou}\ \emph {et~al.}(2024)\citenamefont {Malnou},
  \citenamefont {Larson}, \citenamefont {Teufel}, \citenamefont {Lecocq},\ and\
  \citenamefont {Aumentado}}]{SNT_Malnou2024}%
  \BibitemOpen
  \bibfield  {author} {\bibinfo {author} {\bibfnamefont {M.}~\bibnamefont
  {Malnou}}, \bibinfo {author} {\bibfnamefont {T.~F.~Q.}\ \bibnamefont
  {Larson}}, \bibinfo {author} {\bibfnamefont {J.~D.}\ \bibnamefont {Teufel}},
  \bibinfo {author} {\bibfnamefont {F.}~\bibnamefont {Lecocq}}, \ and\ \bibinfo
  {author} {\bibfnamefont {J.}~\bibnamefont {Aumentado}},\ }\bibfield  {title}
  {\enquote {\bibinfo {title} {{Low-noise cryogenic microwave amplifier
  characterization with a calibrated noise source}},}\ }\href {\doibase
  10.1063/5.0193591} {\bibfield  {journal} {\bibinfo  {journal} {Review of
  Scientific Instruments}\ }\textbf {\bibinfo {volume} {95}},\ \bibinfo {pages}
  {034703} (\bibinfo {year} {2024})}\BibitemShut {NoStop}%
\bibitem [{\citenamefont {Caves}(1982)}]{Cave}%
  \BibitemOpen
  \bibfield  {author} {\bibinfo {author} {\bibfnamefont {Carlton~M.}\
  \bibnamefont {Caves}},\ }\bibfield  {title} {\enquote {\bibinfo {title}
  {Quantum limits on noise in linear amplifiers},}\ }\href {\doibase
  10.1103/PhysRevD.26.1817} {\bibfield  {journal} {\bibinfo  {journal} {Phys.
  Rev. D}\ }\textbf {\bibinfo {volume} {26}},\ \bibinfo {pages} {1817--1839}
  (\bibinfo {year} {1982})}\BibitemShut {NoStop}%
\bibitem [{\citenamefont {Grimm}\ \emph {et~al.}(2020)\citenamefont {Grimm},
  \citenamefont {Frattini}, \citenamefont {Puri}, \citenamefont {Mundhada},
  \citenamefont {Touzard}, \citenamefont {Mirrahimi}, \citenamefont {Girvin},
  \citenamefont {Shankar},\ and\ \citenamefont {Devoret}}]{Kerr_Cat_2020}%
  \BibitemOpen
  \bibfield  {author} {\bibinfo {author} {\bibfnamefont {A.}~\bibnamefont
  {Grimm}}, \bibinfo {author} {\bibfnamefont {N.~E.}\ \bibnamefont {Frattini}},
  \bibinfo {author} {\bibfnamefont {S.}~\bibnamefont {Puri}}, \bibinfo {author}
  {\bibfnamefont {S.~O.}\ \bibnamefont {Mundhada}}, \bibinfo {author}
  {\bibfnamefont {S.}~\bibnamefont {Touzard}}, \bibinfo {author} {\bibfnamefont
  {M.}~\bibnamefont {Mirrahimi}}, \bibinfo {author} {\bibfnamefont {S.~M.}\
  \bibnamefont {Girvin}}, \bibinfo {author} {\bibfnamefont {S.}~\bibnamefont
  {Shankar}}, \ and\ \bibinfo {author} {\bibfnamefont {M.~H.}\ \bibnamefont
  {Devoret}},\ }\bibfield  {title} {\enquote {\bibinfo {title} {Stabilization
  and operation of a {Kerr}-cat qubit},}\ }\href {\doibase
  10.1038/s41586-020-2587-z} {\bibfield  {journal} {\bibinfo  {journal}
  {Nature}\ }\textbf {\bibinfo {volume} {584}},\ \bibinfo {pages} {205--209}
  (\bibinfo {year} {2020})}\BibitemShut {NoStop}%
\bibitem [{\citenamefont {Campagne-Ibarcq}\ \emph {et~al.}(2020)\citenamefont
  {Campagne-Ibarcq}, \citenamefont {Eickbusch}, \citenamefont {Touzard},
  \citenamefont {Zalys-Geller}, \citenamefont {Frattini}, \citenamefont
  {Sivak}, \citenamefont {Reinhold}, \citenamefont {Puri}, \citenamefont
  {Shankar}, \citenamefont {Schoelkopf}, \citenamefont {Frunzio}, \citenamefont
  {Mirrahimi},\ and\ \citenamefont {Devoret}}]{GKP_2020}%
  \BibitemOpen
  \bibfield  {author} {\bibinfo {author} {\bibfnamefont {P.}~\bibnamefont
  {Campagne-Ibarcq}}, \bibinfo {author} {\bibfnamefont {A.}~\bibnamefont
  {Eickbusch}}, \bibinfo {author} {\bibfnamefont {S.}~\bibnamefont {Touzard}},
  \bibinfo {author} {\bibfnamefont {E.}~\bibnamefont {Zalys-Geller}}, \bibinfo
  {author} {\bibfnamefont {N.~E.}\ \bibnamefont {Frattini}}, \bibinfo {author}
  {\bibfnamefont {V.~V.}\ \bibnamefont {Sivak}}, \bibinfo {author}
  {\bibfnamefont {P.}~\bibnamefont {Reinhold}}, \bibinfo {author}
  {\bibfnamefont {S.}~\bibnamefont {Puri}}, \bibinfo {author} {\bibfnamefont
  {S.}~\bibnamefont {Shankar}}, \bibinfo {author} {\bibfnamefont {R.~J.}\
  \bibnamefont {Schoelkopf}}, \bibinfo {author} {\bibfnamefont
  {L.}~\bibnamefont {Frunzio}}, \bibinfo {author} {\bibfnamefont
  {M.}~\bibnamefont {Mirrahimi}}, \ and\ \bibinfo {author} {\bibfnamefont
  {M.~H.}\ \bibnamefont {Devoret}},\ }\bibfield  {title} {\enquote {\bibinfo
  {title} {Quantum error correction of a qubit encoded in grid states of an
  oscillator},}\ }\href {\doibase 10.1038/s41586-020-2603-3} {\bibfield
  {journal} {\bibinfo  {journal} {Nature}\ }\textbf {\bibinfo {volume} {584}},\
  \bibinfo {pages} {368--372} (\bibinfo {year} {2020})}\BibitemShut {NoStop}%
\bibitem [{\citenamefont {Zhou}\ \emph {et~al.}(2023)\citenamefont {Zhou},
  \citenamefont {Lu}, \citenamefont {Praquin}, \citenamefont {Chien},
  \citenamefont {Kaufman}, \citenamefont {Cao}, \citenamefont {Xia},
  \citenamefont {Mong}, \citenamefont {Pfaff}, \citenamefont {Pekker},\ and\
  \citenamefont {Hatridge}}]{zhouRealizingAlltoallCouplings2023}%
  \BibitemOpen
  \bibfield  {author} {\bibinfo {author} {\bibfnamefont {Chao}\ \bibnamefont
  {Zhou}}, \bibinfo {author} {\bibfnamefont {Pinlei}\ \bibnamefont {Lu}},
  \bibinfo {author} {\bibfnamefont {Matthieu}\ \bibnamefont {Praquin}},
  \bibinfo {author} {\bibfnamefont {Tzu-Chiao}\ \bibnamefont {Chien}}, \bibinfo
  {author} {\bibfnamefont {Ryan}\ \bibnamefont {Kaufman}}, \bibinfo {author}
  {\bibfnamefont {Xi}~\bibnamefont {Cao}}, \bibinfo {author} {\bibfnamefont
  {Mingkang}\ \bibnamefont {Xia}}, \bibinfo {author} {\bibfnamefont {Roger
  S.~K.}\ \bibnamefont {Mong}}, \bibinfo {author} {\bibfnamefont {Wolfgang}\
  \bibnamefont {Pfaff}}, \bibinfo {author} {\bibfnamefont {David}\ \bibnamefont
  {Pekker}}, \ and\ \bibinfo {author} {\bibfnamefont {Michael}\ \bibnamefont
  {Hatridge}},\ }\bibfield  {title} {\enquote {\bibinfo {title} {Realizing
  all-to-all couplings among detachable quantum modules using a microwave
  quantum state router},}\ }\href {\doibase 10.1038/s41534-023-00723-7}
  {\bibfield  {journal} {\bibinfo  {journal} {npj Quantum Information}\
  }\textbf {\bibinfo {volume} {9}},\ \bibinfo {pages} {1--9} (\bibinfo {year}
  {2023})}\BibitemShut {NoStop}%
\bibitem [{\citenamefont {Chapman}\ \emph {et~al.}(2023)\citenamefont
  {Chapman}, \citenamefont {De~Graaf}, \citenamefont {Xue}, \citenamefont
  {Zhang}, \citenamefont {Teoh}, \citenamefont {Curtis}, \citenamefont
  {Tsunoda}, \citenamefont {Eickbusch}, \citenamefont {Read}, \citenamefont
  {Koottandavida}, \citenamefont {Mundhada}, \citenamefont {Frunzio},
  \citenamefont {Devoret}, \citenamefont {Girvin},\ and\ \citenamefont
  {Schoelkopf}}]{chapmanHighOnOffRatioBeamSplitterInteraction2023}%
  \BibitemOpen
  \bibfield  {author} {\bibinfo {author} {\bibfnamefont {Benjamin~J.}\
  \bibnamefont {Chapman}}, \bibinfo {author} {\bibfnamefont {Stijn~J.}\
  \bibnamefont {De~Graaf}}, \bibinfo {author} {\bibfnamefont {Sophia~H.}\
  \bibnamefont {Xue}}, \bibinfo {author} {\bibfnamefont {Yaxing}\ \bibnamefont
  {Zhang}}, \bibinfo {author} {\bibfnamefont {James}\ \bibnamefont {Teoh}},
  \bibinfo {author} {\bibfnamefont {Jacob~C.}\ \bibnamefont {Curtis}}, \bibinfo
  {author} {\bibfnamefont {Takahiro}\ \bibnamefont {Tsunoda}}, \bibinfo
  {author} {\bibfnamefont {Alec}\ \bibnamefont {Eickbusch}}, \bibinfo {author}
  {\bibfnamefont {Alexander~P.}\ \bibnamefont {Read}}, \bibinfo {author}
  {\bibfnamefont {Akshay}\ \bibnamefont {Koottandavida}}, \bibinfo {author}
  {\bibfnamefont {Shantanu~O.}\ \bibnamefont {Mundhada}}, \bibinfo {author}
  {\bibfnamefont {Luigi}\ \bibnamefont {Frunzio}}, \bibinfo {author}
  {\bibfnamefont {M.H.}\ \bibnamefont {Devoret}}, \bibinfo {author}
  {\bibfnamefont {S.M.}\ \bibnamefont {Girvin}}, \ and\ \bibinfo {author}
  {\bibfnamefont {R.J.}\ \bibnamefont {Schoelkopf}},\ }\bibfield  {title}
  {\enquote {\bibinfo {title} {High-{On}-{Off}-{Ratio} {Beam}-{Splitter}
  {Interaction} for {Gates} on {Bosonically} {Encoded} {Qubits}},}\ }\href
  {\doibase 10.1103/PRXQuantum.4.020355} {\bibfield  {journal} {\bibinfo
  {journal} {PRX Quantum}\ }\textbf {\bibinfo {volume} {4}},\ \bibinfo {pages}
  {020355} (\bibinfo {year} {2023})}\BibitemShut {NoStop}%
\bibitem [{\citenamefont {Xia}\ \emph {et~al.}(2023)\citenamefont {Xia},
  \citenamefont {Zhou}, \citenamefont {Liu}, \citenamefont {Patel},
  \citenamefont {Cao}, \citenamefont {Lu}, \citenamefont {Mesits},
  \citenamefont {Mucci}, \citenamefont {Gorski}, \citenamefont {Pekker},\ and\
  \citenamefont {Hatridge}}]{xiaFastSuperconductingQubit2023}%
  \BibitemOpen
  \bibfield  {author} {\bibinfo {author} {\bibfnamefont {Mingkang}\
  \bibnamefont {Xia}}, \bibinfo {author} {\bibfnamefont {Chao}\ \bibnamefont
  {Zhou}}, \bibinfo {author} {\bibfnamefont {Chenxu}\ \bibnamefont {Liu}},
  \bibinfo {author} {\bibfnamefont {Param}\ \bibnamefont {Patel}}, \bibinfo
  {author} {\bibfnamefont {Xi}~\bibnamefont {Cao}}, \bibinfo {author}
  {\bibfnamefont {Pinlei}\ \bibnamefont {Lu}}, \bibinfo {author} {\bibfnamefont
  {Boris}\ \bibnamefont {Mesits}}, \bibinfo {author} {\bibfnamefont {Maria}\
  \bibnamefont {Mucci}}, \bibinfo {author} {\bibfnamefont {David}\ \bibnamefont
  {Gorski}}, \bibinfo {author} {\bibfnamefont {David}\ \bibnamefont {Pekker}},
  \ and\ \bibinfo {author} {\bibfnamefont {Michael}\ \bibnamefont {Hatridge}},\
  }\href@noop {} {\enquote {\bibinfo {title} {Fast superconducting qubit
  control with sub-harmonic drives},}\ } (\bibinfo {year} {2023}),\ \bibinfo
  {note} {\url{https://arxiv.org/abs/2306.10162}}\BibitemShut {NoStop}%
\bibitem [{\citenamefont {Xu}\ \emph {et~al.}(2024)\citenamefont {Xu},
  \citenamefont {Wu}, \citenamefont {Dai},\ and\ \citenamefont
  {Tang}}]{xuRadiativelyCooledQuantum2024}%
  \BibitemOpen
  \bibfield  {author} {\bibinfo {author} {\bibfnamefont {Mingrui}\ \bibnamefont
  {Xu}}, \bibinfo {author} {\bibfnamefont {Yufeng}\ \bibnamefont {Wu}},
  \bibinfo {author} {\bibfnamefont {Wei}\ \bibnamefont {Dai}}, \ and\ \bibinfo
  {author} {\bibfnamefont {Hong~X.}\ \bibnamefont {Tang}},\ }\bibfield  {title}
  {\enquote {\bibinfo {title} {Radiatively cooled quantum microwave
  amplifiers},}\ }\href
  {https://pubs.aip.org/aip/apl/article/125/2/024001/3302658} {\bibfield
  {journal} {\bibinfo  {journal} {Applied Physics Letters}\ }\textbf {\bibinfo
  {volume} {125}},\ \bibinfo {pages} {024001} (\bibinfo {year}
  {2024})}\BibitemShut {NoStop}%
\bibitem [{\citenamefont {Spietz}\ \emph {et~al.}(2003)\citenamefont {Spietz},
  \citenamefont {Lehnert}, \citenamefont {Siddiqi},\ and\ \citenamefont
  {Schoelkopf}}]{SNT}%
  \BibitemOpen
  \bibfield  {author} {\bibinfo {author} {\bibfnamefont {Lafe}\ \bibnamefont
  {Spietz}}, \bibinfo {author} {\bibfnamefont {K.~W.}\ \bibnamefont {Lehnert}},
  \bibinfo {author} {\bibfnamefont {I.}~\bibnamefont {Siddiqi}}, \ and\
  \bibinfo {author} {\bibfnamefont {R.~J.}\ \bibnamefont {Schoelkopf}},\
  }\bibfield  {title} {\enquote {\bibinfo {title} {Primary electronic
  thermometry using the shot noise of a tunnel junction},}\ }\href {\doibase
  10.1126/science.1084647} {\bibfield  {journal} {\bibinfo  {journal}
  {Science}\ }\textbf {\bibinfo {volume} {300}},\ \bibinfo {pages} {1929--1932}
  (\bibinfo {year} {2003})}\BibitemShut {NoStop}%
\bibitem [{\citenamefont {Minev}\ \emph {et~al.}(2021)\citenamefont {Minev},
  \citenamefont {Leghtas}, \citenamefont {Mundhada}, \citenamefont
  {Christakis}, \citenamefont {Pop},\ and\ \citenamefont
  {Devoret}}]{minevEnergyparticipationQuantizationJosephson2021}%
  \BibitemOpen
  \bibfield  {author} {\bibinfo {author} {\bibfnamefont {Zlatko~K.}\
  \bibnamefont {Minev}}, \bibinfo {author} {\bibfnamefont {Zaki}\ \bibnamefont
  {Leghtas}}, \bibinfo {author} {\bibfnamefont {Shantanu~O.}\ \bibnamefont
  {Mundhada}}, \bibinfo {author} {\bibfnamefont {Lysander}\ \bibnamefont
  {Christakis}}, \bibinfo {author} {\bibfnamefont {Ioan~M.}\ \bibnamefont
  {Pop}}, \ and\ \bibinfo {author} {\bibfnamefont {Michel~H.}\ \bibnamefont
  {Devoret}},\ }\bibfield  {title} {\enquote {\bibinfo {title}
  {Energy-participation quantization of {Josephson} circuits},}\ }\href
  {\doibase 10.1038/s41534-021-00461-8} {\bibfield  {journal} {\bibinfo
  {journal} {npj Quantum Information}\ }\textbf {\bibinfo {volume} {7}},\
  \bibinfo {pages} {1--11} (\bibinfo {year} {2021})}\BibitemShut {NoStop}%
\bibitem [{\citenamefont {Oppenheim}\ \emph {et~al.}(2010)\citenamefont
  {Oppenheim}, \citenamefont {Willsky},\ and\ \citenamefont
  {Nawab}}]{oppenheim_ss}%
  \BibitemOpen
  \bibfield  {author} {\bibinfo {author} {\bibfnamefont {A.V.}\ \bibnamefont
  {Oppenheim}}, \bibinfo {author} {\bibfnamefont {A.S.}\ \bibnamefont
  {Willsky}}, \ and\ \bibinfo {author} {\bibfnamefont {S.H.}\ \bibnamefont
  {Nawab}},\ }\href@noop {} {\emph {\bibinfo {title} {Signals and Systems}}}\
  (\bibinfo  {publisher} {Prentice-Hall Inc.},\ \bibinfo {address} {Hoboken,
  NJ},\ \bibinfo {year} {2010})\BibitemShut {NoStop}%
\bibitem [{\citenamefont {Nigg}\ \emph {et~al.}(2012)\citenamefont {Nigg},
  \citenamefont {Paik}, \citenamefont {Vlastakis}, \citenamefont {Kirchmair},
  \citenamefont {Shankar}, \citenamefont {Frunzio}, \citenamefont {Devoret},
  \citenamefont {Schoelkopf},\ and\ \citenamefont
  {Girvin}}]{niggBlackBoxSuperconductingCircuit2012}%
  \BibitemOpen
  \bibfield  {author} {\bibinfo {author} {\bibfnamefont {Simon~E.}\
  \bibnamefont {Nigg}}, \bibinfo {author} {\bibfnamefont {Hanhee}\ \bibnamefont
  {Paik}}, \bibinfo {author} {\bibfnamefont {Brian}\ \bibnamefont {Vlastakis}},
  \bibinfo {author} {\bibfnamefont {Gerhard}\ \bibnamefont {Kirchmair}},
  \bibinfo {author} {\bibfnamefont {S.}~\bibnamefont {Shankar}}, \bibinfo
  {author} {\bibfnamefont {Luigi}\ \bibnamefont {Frunzio}}, \bibinfo {author}
  {\bibfnamefont {M.~H.}\ \bibnamefont {Devoret}}, \bibinfo {author}
  {\bibfnamefont {R.~J.}\ \bibnamefont {Schoelkopf}}, \ and\ \bibinfo {author}
  {\bibfnamefont {S.~M.}\ \bibnamefont {Girvin}},\ }\bibfield  {title}
  {\enquote {\bibinfo {title} {Black-box superconducting circuit
  quantization},}\ }\href {\doibase 10.1103/PhysRevLett.108.240502} {\bibfield
  {journal} {\bibinfo  {journal} {Phys. Rev. Lett.}\ }\textbf {\bibinfo
  {volume} {108}},\ \bibinfo {pages} {240502} (\bibinfo {year}
  {2012})}\BibitemShut {NoStop}%
\bibitem [{\citenamefont {Gao}\ \emph {et~al.}(2019)\citenamefont {Gao},
  \citenamefont {Lester}, \citenamefont {Chou}, \citenamefont {Frunzio},
  \citenamefont {Devoret}, \citenamefont {Jiang}, \citenamefont {Girvin},\ and\
  \citenamefont {Schoelkopf}}]{gaoEntanglementBosonicModes2019}%
  \BibitemOpen
  \bibfield  {author} {\bibinfo {author} {\bibfnamefont {Yvonne~Y.}\
  \bibnamefont {Gao}}, \bibinfo {author} {\bibfnamefont {Brian~J.}\
  \bibnamefont {Lester}}, \bibinfo {author} {\bibfnamefont {Kevin~S.}\
  \bibnamefont {Chou}}, \bibinfo {author} {\bibfnamefont {Luigi}\ \bibnamefont
  {Frunzio}}, \bibinfo {author} {\bibfnamefont {Michel~H.}\ \bibnamefont
  {Devoret}}, \bibinfo {author} {\bibfnamefont {Liang}\ \bibnamefont {Jiang}},
  \bibinfo {author} {\bibfnamefont {S.~M.}\ \bibnamefont {Girvin}}, \ and\
  \bibinfo {author} {\bibfnamefont {Robert~J.}\ \bibnamefont {Schoelkopf}},\
  }\bibfield  {title} {\enquote {\bibinfo {title} {Entanglement of bosonic
  modes through an engineered exchange interaction},}\ }\href {\doibase
  10.1038/s41586-019-0970-4} {\bibfield  {journal} {\bibinfo  {journal}
  {Nature}\ }\textbf {\bibinfo {volume} {566}},\ \bibinfo {pages} {509--512}
  (\bibinfo {year} {2019})}\BibitemShut {NoStop}%
\bibitem [{\citenamefont {Ding}\ \emph {et~al.}(2024)\citenamefont {Ding},
  \citenamefont {Brock}, \citenamefont {Eickbusch}, \citenamefont
  {Koottandavida}, \citenamefont {Frattini}, \citenamefont {Cortinas},
  \citenamefont {Joshi}, \citenamefont {de~Graaf}, \citenamefont {Chapman},
  \citenamefont {Ganjam}, \citenamefont {Frunzio}, \citenamefont {Schoelkopf},\
  and\ \citenamefont {Devoret}}]{dingQuantumControlOscillator2024}%
  \BibitemOpen
  \bibfield  {author} {\bibinfo {author} {\bibfnamefont {Andy~Z.}\ \bibnamefont
  {Ding}}, \bibinfo {author} {\bibfnamefont {Benjamin~L.}\ \bibnamefont
  {Brock}}, \bibinfo {author} {\bibfnamefont {Alec}\ \bibnamefont {Eickbusch}},
  \bibinfo {author} {\bibfnamefont {Akshay}\ \bibnamefont {Koottandavida}},
  \bibinfo {author} {\bibfnamefont {Nicholas~E.}\ \bibnamefont {Frattini}},
  \bibinfo {author} {\bibfnamefont {Rodrigo~G.}\ \bibnamefont {Cortinas}},
  \bibinfo {author} {\bibfnamefont {Vidul~R.}\ \bibnamefont {Joshi}}, \bibinfo
  {author} {\bibfnamefont {Stijn~J.}\ \bibnamefont {de~Graaf}}, \bibinfo
  {author} {\bibfnamefont {Benjamin~J.}\ \bibnamefont {Chapman}}, \bibinfo
  {author} {\bibfnamefont {Suhas}\ \bibnamefont {Ganjam}}, \bibinfo {author}
  {\bibfnamefont {Luigi}\ \bibnamefont {Frunzio}}, \bibinfo {author}
  {\bibfnamefont {Robert~J.}\ \bibnamefont {Schoelkopf}}, \ and\ \bibinfo
  {author} {\bibfnamefont {Michel~H.}\ \bibnamefont {Devoret}},\ }\href@noop {}
  {\enquote {\bibinfo {title} {Quantum control of an oscillator with a kerr-cat
  qubit},}\ } (\bibinfo {year} {2024}),\ \bibinfo {note}
  {\url{https://arxiv.org/abs/2407.10940}}\BibitemShut {NoStop}%
\bibitem [{\citenamefont {Miano}\ \emph {et~al.}(2022)\citenamefont {Miano},
  \citenamefont {Liu}, \citenamefont {Sivak}, \citenamefont {Frattini},
  \citenamefont {Joshi}, \citenamefont {Dai}, \citenamefont {Frunzio},\ and\
  \citenamefont {Devoret}}]{G-SNAIL}%
  \BibitemOpen
  \bibfield  {author} {\bibinfo {author} {\bibfnamefont {A.}~\bibnamefont
  {Miano}}, \bibinfo {author} {\bibfnamefont {G.}~\bibnamefont {Liu}}, \bibinfo
  {author} {\bibfnamefont {V.~V.}\ \bibnamefont {Sivak}}, \bibinfo {author}
  {\bibfnamefont {N.~E.}\ \bibnamefont {Frattini}}, \bibinfo {author}
  {\bibfnamefont {V.~R.}\ \bibnamefont {Joshi}}, \bibinfo {author}
  {\bibfnamefont {W.}~\bibnamefont {Dai}}, \bibinfo {author} {\bibfnamefont
  {L.}~\bibnamefont {Frunzio}}, \ and\ \bibinfo {author} {\bibfnamefont
  {M.~H.}\ \bibnamefont {Devoret}},\ }\bibfield  {title} {\enquote {\bibinfo
  {title} {Frequency-tunable {{Kerr-free}} three-wave mixing with a
  gradiometric {{SNAIL}}},}\ }\href {\doibase 10.1063/5.0083350} {\bibfield
  {journal} {\bibinfo  {journal} {Applied Physics Letters}\ }\textbf {\bibinfo
  {volume} {120}},\ \bibinfo {pages} {184002} (\bibinfo {year}
  {2022})}\BibitemShut {NoStop}%
\end{thebibliography}

\end{document}